\documentclass[preprint,preprintnumbers,amsmath,amssymb]{revtex4}
\usepackage{graphicx}

\begin{document}

\title{Laws of Population Growth}

\author{Hern\'an D. Rozenfeld$^{1}$, Diego Rybski$^{1}$,
Jos\'e S. Andrade Jr.$^{2}$,
Michael Batty$^{3}$, H. Eugene Stanley$^{4}$,
and Hern\'an A. Makse$^{1,2}$}

\affiliation{$^1$Levich Institute and Physics Department, City College
  of New York,
  New York, NY 10031, USA\\
  $^2$Departamento de F\'\i sica, Universidade Federal do
  Cear\'a, 60451-970 Fortaleza, Cear\'a, Brazil\\
  $^3$Centre for Advanced Spatial Analysis, University College
  London,\\ 1-19 Torrington Place, London WC1E 6BT, UK\\ $^4$Center
  for Polymer Studies and Physics Department, Boston University,
  Boston, MA 02215, USA }

\begin{abstract}

  An important issue in the study of cities is defining a metropolitan
  area, as different definitions affect conclusions regarding the statistical distribution
  of urban activity.  A commonly employed method of defining a
  metropolitan area is the Metropolitan Statistical Areas (MSAs), based
  on rules attempting to capture the notion of city as a functional
  economic region, and is performed using experience. The construction of MSAs is a
  time-consuming process and is typically done only for a subset (a few
  hundreds) of the most highly populated cities.  Here, we introduce a
  new method to designate metropolitan areas, denoted ``City
  Clustering Algorithm'' (CCA).  The CCA is based on spatial
  distributions of the population at a fine geographic scale, defining
  a city beyond the scope of its administrative boundaries.  We use
  the CCA to examine Gibrat's law of proportional growth, which postulates 
  that the mean and standard deviation of the growth rate of cities
  are constant, independent of city size.  We find that the mean
  growth rate of a cluster utilizing the CCA exhibits deviations from
  Gibrat's law, and that the standard deviation decreases as a
  power-law with respect to the city size. The CCA allows for the
  study of the underlying process leading to these deviations, which are shown
  to arise from the existence of long-range spatial correlations in
  population growth.  These results have socio-political
  implications, for example for the location of new
  economic development in cities of varied size.

\end{abstract}

\maketitle

\section{Introduction}

In recent years there has been considerable work on how to define
cities and how the different definitions affect
the statistical distribution of urban activity~\cite{Gabaix99,Gab03}.
This is a long standing problem in spatial analysis of aggregated data
sources, referred to as the `modifiable areal unit problem' or the
`ecological fallacy'~\cite{unwin,king},
where different definitions of spatial units based on administrative
or governmental boundaries, give rise to inconsistent conclusions with
respect to explanations and interpretations of data at different
scales. The conventional method of defining human agglomerations is
through the MSAs~\cite{Gabaix99,Gab03,dobkins00,Ioa03,Eeckhout04},
which is subject to socio-economical factors. The MSA has been of
indubitable importance for the analysis of population growth,
and is constructed manually case-by-case based on subjective judgment (MSAs are defined starting from a highly populated central area and adding its surrounding counties if they have social or economical ties).

In this report, we propose a new way to measure the extent of human
agglomerations based on clustering techniques using a fine geographical
grid, covering both urban and rural areas. In this view, ``cities''
represent clusters of population, i.e., adjacent populated
geo\-graphical spaces. Our algorithm, the ``city clustering algorithm''
(CCA), allows for an automated and systematic way of building
population clusters based on the geographical location of people. The
CCA has one parameter (the cell size) that is useful for the study of
human agglomerations at different length scales, similar to the level
of aggregation in the context of social sciences. We show that the CCA
allows for the study of the origin of statistical properties of
population growth.  We use the CCA to analyze the postulates of
Gibrat's law of proportional growth applied to cities, which assumes
that the mean and standard deviation of the growth rates of cities are
constant.  We show that population growth at a fine geographical scale
for different urban and regional systems at country and continental
levels (Great Britain, the USA, and Africa) deviates from Gibrat's
law. We find that the mean and standard deviation of population growth
rates decrease with population size, in some cases following a
power-law behavior.  We argue that the underlying demographic process
leading to the deviations from Gibrat's law can be modeled from the
existence of long-range spatial correlations in the growth of the
population, which may arise from the concept that ``development
attracts further development.''  These results have implications for
social policies, such as those pertaining to the location of new
economic development in cities of different sizes.  The present
results imply that, on average, the greatest growth rate occurs in the
smallest places where there is the greatest risk of failure (larger
fluctuations).  A corollary is that the safest growth occurs in the
largest places having less likelihood for rapid growth.

The analyzed data consist of the number of inhabitants, $n_i(t)$, in
each cell $i$ of a fine geographical grid at a given time, $t$.  The
cell size varies for each data set used in this study.  We consider
three different geographic scales: on the smallest scale, the area of
study is Great Britain (GB: England, Scotland and Wales), a highly
urbanized country with population of 58.7 million in 2007, and an area
of 0.23 million km$^2$. The grid is composed of 5.75 million cells of
200m-by-200m~\cite{Census}. At the intermediate scale, we study the
USA (continental USA without Alaska), a single country nearly
continental in scale, with a population of 303 million in 2007, and an
area of 7.44 million km$^2$. The grid contains 7.44 million cells of
approximately 1km-by-1km obtained from the US Census Bureau
\cite{ESRI}.  The datasets of GB and USA are populated-places
datasets, with population counts defined at points in a grid.
Since there could be some distortions in the true residential
population involved at the finest grid resolution, we perform our
analysis by investigating the statistical properties as a function of
the grid size by coarse-graining the data as explained in Section
\ref{details}.  At the largest scale, we analyze the continent of
Africa, composed of 53 countries with a total population of 933
million in 2007, and an area of 30.34 million km$^2$. These data are
gridded with less resolution by 0.50 million cells of approximately
7.74km-by-7.74km~\cite{UNESCO}.  More detailed information about these
datasets is found in Section~\ref{details} (all the datasets studied
in this paper are available at
http://lev.ccny.cuny.edu/$\sim$hmakse/cities/city\_data.zip).

\section{Results}

Figure \ref{fig1}A illustrates operation of the CCA. In order to
identify urban clusters, we require connected cells to have nonzero
population. We start by selecting an arbitrary populated cell (final results are independent of the choice of the initial cell). Iteratively,
we then grow a cluster by adding nearest neighbors of the boundary cells
with a population strictly greater than zero, until all neighbors of the boundary are
unpopulated. We repeat this process until all populated cells have
been assigned to a cluster.  This technique was introduced to model
forest fire dynamics \cite{Stauffer84} and is termed the ``burning
algorithm,'' since one can think of each populated cell as a burning
tree.

The population $S_i(t)$ of cluster $i$ at time $t$ is the sum of the
populations $n_{j}^{(i)}(t)$ of each cell $j$ within it, $S_i(t) =
\sum_{j=1}^{N_i} n_j^{(i)}(t)$, where $N_i$ is the number of cells in
the cluster. Results of the CCA are shown in Fig.~\ref{fig1}B,
representing the urban cluster surrounding the City of London (red
cluster overlaying a satellite image,
see http://lev.ccny.cuny.edu/$\sim$hmakse/cities/london.gif for an
animated image of Fig.~\ref{fig1}B). Figure \ref{fig1}C 
depicts all the clusters of GB, indicating the large
variability in their population and size.

A feature of the CCA is that it allows the analysis of the population
clusters at different length scales
by coarse-graining the grid
and applying the CCA to the coarse-grained dataset (see
Section~\ref{details} for details on coarse-graining the data).
At larger scales, disconnected areas around the edge of a cluster
could be added into the cluster.  This is justified when, for example,
a town is divided by a wide highway or a river.

Tables~I and~II in Supporting Information (SI)
Section~I. show a detailed comparison between the urban
clusters obtained with the CCA applied to the USA in 1990, and the
results obtained from the analysis of MSAs from the US Census Bureau used
in previous studies of population growth
~\cite{dobkins00,Ioa03,Eeckhout04}.  We observe that the MSAs
considered in Ref.~\cite{Eeckhout04} are similar to the clusters
obtained with the CCA with a cell size of 4km-by-4km or 8km-by-8km.
In particular, the population sizes of the clusters have the same
order of magnitude as the MSAs. On the other hand, for large cities the
MSAs from the data of Ref.~\cite{dobkins00} seem to be mostly
comparable to our results for cell sizes of 2km-by-2km or 4km-by-4km.

Use of the CCA permits a systematic study of
cluster dynamics. For instance, clusters may expand or contract, merge
or split between two considered times as illustrated in
Fig.~\ref{cases}. We quantify these processes by measuring the
probability distribution of the temporal changes in the clusters for
the data of GB.
We find that when the cell size is 2.2km-by-2.2km, 84\% of the
clusters evolve from 1981 to 1991 following the three first cases
presented in Fig.~\ref{cases} (no change, expansion or reduction), 6\%
of the clusters merge from two clusters into one in 1991, and 3\% of
the clusters split into two clusters.

Next, we apply the CCA to study the dynamics of population growth by
investigating Gibrat's law, which postulates that the mean and
standard deviation of growth rates are constant
\cite{Eeckhout04,Gabaix99,Ioa03,Gab03,Eaton97}.  The conventional
method~\cite{Gabaix99,Ioa03,Gab03} is to assume that the populations of a
given city or cluster $i$, at times $t_0$ and $t_1>t_0$, are related
by
\begin{equation}
S_1=R(S_0)S_0,
\label{eq:growth}
\end{equation}
where $S_{0} \equiv S_i(t_0) = \sum_{j}^{N_i} n_j^{(i)}(t_0)$ and
$S_{1} \equiv S_i(t_1) = \sum_{j}^{N_i} n_j^{(i)}(t_1)$ are the
initial and final populations of cluster $i$, respectively, and
$R(S_{0})$ is the positive growth factor which varies from cluster to
cluster. Following the literature in population dynamics~\cite{Gabaix99,Ioa03,Gab03,Eeckhout04}, we define the population growth rate of a cluster as
$r(S_0)\equiv\ln R(S_0) =\ln(S_{1}/S_{0})$, and
study the dependence of the mean value of the growth rate, $\langle
r(S_{0}) \rangle$, and the standard deviation,
$\sigma(S_{0})=\sqrt{\langle r(S_{0})^{2}\rangle -\langle
  r(S_{0})\rangle^{2}}$, on the initial population, $S_0$. The
averages $\langle r(S_{0}) \rangle$ and $\sigma(S_{0})$ are calculated
applying nonparametric techniques ~\cite{hardle, Silverman86} (see
Section~\ref{expla} for details). To obtain the population growth rate
of clusters we take into account that not all clusters occupy the same
area between $t_0$ and $t_1$ according to the cases discussed in
Fig. \ref{cases}. The figure shows how to calculate the growth rate
$r(S_0)$ in each case.

We analyze the population growth in the USA from $t_0=1990$ to $t_1=2000$
\cite{ESRI}. We apply the CCA to identify the clusters in the data of
1990 and calculate their growth rates by comparing them to the population of
the same clusters in 2000 when the data are gridded with a cell size of
2000m by 2000m. We calculate the annual growth rates by dividing $r$
by the time interval $t_1-t_0$.

Figure~\ref{usascaling}A shows a nonparametric regression with
bootstrapped 95\% confidence bands~\cite{hardle, Silverman86} of the
growth rate of the USA, $\langle r(S_0)\rangle$ (see 
Section~\ref{expla} for details).  We find that the growth rate
diminishes from $\langle r(S_0) \rangle \approx 0.012\pm 0.004$ (error
includes the confidence bands) for populations below $10^4$
inhabitants to $\langle r(S_0) \rangle \approx 0.002\pm 0.002$ for the
largest populations around $S_0 \approx 10^7$.  We may argue that the
mean growth rate deviates from Gibrat's law beyond the confidence
bands.  While it is difficult to fit the data to a single function for
the entire range, the data show a decrease with $S_0$ approximately
following a power-law in the tail for populations larger than
$10^4$. An attempt to fit the data with a power-law yields the
following scaling in the tail:
\begin{equation}
  \langle r(S_0)
  \rangle \sim S_0^{-\alpha},
\end{equation}
where $\alpha$ is the mean growth exponent, that takes a value
$\alpha_{\rm USA}=0.28\pm0.08$ from Ordinary Least Squares (OLS)
analysis~\cite{montgomery} (see Section~\ref{expla} for details on
OLS and on the estimation of the exponent error).

Figure~\ref{usascaling}B shows the dependence of the standard
deviation $\sigma(S_0)$ on the initial population $S_0$.  On average,
fluctuations in the growth rate of large cities are smaller than for small
cities in contrast to Gibrat's law.  This result can be approximated
over many orders of magnitude by the power-law,
\begin{equation}
\sigma(S_{0}) \sim S_{0}^{-\beta},
\label{scaling}
\end{equation}
where $\beta$ is the standard deviation exponent. We carry out an OLS
regression analysis and find that $ \beta_{\rm USA} =
0.20\pm0.06$.  The presence of a power-law implies that fluctuations in the growth process are statistically
self-similar at different scales, for populations ranging from
$\sim$1000 to $\sim$10 million according to Fig.~\ref{usascaling}B.

Figure~\ref{ukscaling} shows the analysis of the growth rate of the
population clusters of GB from gridded databases~\cite{Census} with a
cell size of 2.2km-by-2.2km at $t_0= 1981$ and $t_1= 1991$.
The average growth rate depicted in Fig.~\ref{ukscaling}A comprises 
large fluctuations as a function of $S_0$, especially for smaller
populations.  However, a slight decrease with population seems evident
from rates around $\langle r \rangle \approx 0.008\pm0.001$ with $S_0
\approx 10^4$ dropping to zero or even negative values for the largest
populations, $S_0\approx 10^6$.
We find that 3556 clusters with population around $S_0=10^3$ exhibit
negative growth rates as well. Thus, the mean rates are plotted on a
semi-logarithmic scale in Fig.~\ref{ukscaling}A.  
When considering intermediate
populations ranging from $S_0 = 3000$ to $S_0=3 \times 10^5$, the data
seem to be following approximately a power-law with $\alpha_{\rm
  GB}=0.17\pm0.05$ from OLS regression analysis, as shown in the inset
of Fig.~\ref{ukscaling}A.  Figure~\ref{ukscaling}B shows the standard
deviation for GB, $\sigma(S_0)$, exhibiting deviations from Gibrat's
law having a tendency to decrease with population according to
Eq.~(\ref{scaling}) and a standard deviation exponent, $\beta_{\rm GB}
= 0.27\pm0.04$, obtained with OLS technique. 

The CCA allows for a study of the growth rates as a function of the
scale of observation, by changing the size of the grid. 
We find (SI Section~II.) that the data for GB are
approximately invariant under coarse-graining the grid at different
levels for both the mean and standard deviation.  When the data of the USA
are aggregated spatially from cell size 2000m to 8000m, the scaling of
the mean rates crosses-over to a flat behavior closer to Gibrat's
law. At the scale of 8000m the mean is approximately constant (with
fluctuations). However, we find that, at this scale, 
all cities in the northeastern the USA spanning from Boston to Washington D.C. 
form a single cluster.  Despite these differences, the scaling
of the standard deviation for the USA holds approximately invariant even
up to the large scale of observation of 8000m.

Next, we analyze the population growth in Africa during the period
1960 to 1990 \cite{UNESCO}. In this case the population data are based
on a larger cell size, so we evaluate the data cell by cell (without the application of the CCA). Despite
the differences in the economic and urban development of Africa, Great
Britain and the USA,
we find that the mean and standard deviation of the growth rate in
Africa display similar scaling as found for the USA and GB. In
Fig.~\ref{africascaling}A we show the results for the growth rate in
Africa when the grid is coarse-grained with a cell size of 77km-by-77km. We find a decrease of the growth rate from $\langle r(S_0)
\rangle \approx 0.1$ to $\langle r(S_0) \rangle \approx 0.01$ between populations
$S_0 \approx 10^3$ and $S_0 \approx 10^6$, respectively.  
All populations have positive growth rates. A log-log plot of the mean rates
shown in Fig.~\ref{africascaling}A reveals a power-law
scaling $\langle r(S_0) \rangle \sim S_0^{-\alpha_{\rm Af}}$, with
$\alpha_{\rm Af}=0.21\pm0.05$ from OLS regression analysis.  The
standard deviation (Fig.~\ref{africascaling}B) satisfies Eq.~(\ref{scaling}) with a standard
deviation exponent $\beta_{\rm Af} = 0.19 \pm 0.04$.
The CCA allows for a study of the origin of the observed behavior of
the growth rates by examining the dynamics and spatial correlations of the
population of cells. To this end, we first generate a surrogate
dataset that consists of shuffling two randomly chosen populated
cells, $n_j^{(i)}(t_0)$ and $n_k^{(i)}(t_0)$, at time $t_0$. This
swapping process preserves the probability distribution of
$n_j^{(i)}$, but destroys any spatial correlations among the
population cells.  Figure~\ref{ukscaling}C shows the results of the
randomization of the GB dataset, indicating power-law scaling in the
tail of $\sigma(S_0)$ with standard deviation exponent $\beta_{\rm
  rand} = 1/2$. This result can be interpreted in terms of the
uncorrelated nature of the randomized dataset (SI Section~III). We consider that the
population of each cell $j$ increases by a random amount $\delta_j$
with mean value $\bar \delta $ and variance $\langle (\delta -\bar
{\delta})^2\rangle = \Delta^2$, and that $r \ll 1$, then
$n_j^{(i)}(t_1) = n_j^{(i)}(t_0) + \delta_j$. Therefore, the
population of a cluster at time $t_1$ can be written as 
\begin{equation}
S_1 = S_0 +
\sum_{j=1}^{N_i}\delta_j.
\end{equation}
It can be shown that (SI Section~III.):

\begin{equation}
\langle S_{1}^{2} \rangle = \langle S_{0}^{2} \rangle +
\sum_{j}^{N_i} \sum_{k}^{N_i} \langle (\delta_{j}-\bar {\delta})
(\delta_{k}-\bar {\delta}) \rangle.
\end{equation}
Randomly shuffling population cells destroys the correlations,
leading to $\langle (\delta_{j}-\bar {\delta}) (\delta_{k}-\bar
{\delta}) \rangle = \Delta^2 \delta_{jk}$ (where $\delta_{jk}$ is the
Kronecker delta function) which implies $\beta_{\rm rand}=1/2$
\cite{Stanley96} (see SI Section~III.).

The fact that $\beta$ lies below the random exponent ($\beta_{\rm
  rand} =1/2$) for all the analyzed data suggests that the dynamics of
the population cells display spatial correlations, which are
eliminated in the random surrogate data. The cells are not occupied
randomly but spatial correlations arise, since when the population in
one cell increases, the probability of growth in an adjacent cell also
increases. That is, development attracts further development, an idea
that has been used to model the spatial distribution of urban patterns
\cite{Makse95}. Indeed this ideas are related to the study of the origin
of power-laws in complex systems~\cite{barabasi,doyle}.

When we analyze the populated cells, we indeed find that spatial correlations
in the incremental population of the cells, $\delta_j$, are
asymptotically of a scale-invariant form characterized by a
correlation exponent $\gamma$,
\begin{equation}
  \langle
  (\delta_{j}-\bar{\delta}) (\delta_{k}-\bar{\delta}) \rangle \sim \frac{\Delta^2}
  {|\vec{x}_{j} - \vec{x}_{k}|^{\gamma}},
\label{cell}
\end{equation}
where $\vec{x}_j$ is the location of cell $j$. For GB we find $\gamma
= 0.93 \pm 0.08 $ (see Fig.~\ref{ukscaling}D). In SI
Section~III. we show that power-law correlations
in the fluctuations at the cell level, Eq.  (\ref{cell}), lead to a
standard deviation exponent $\beta = \gamma/4$. For $\gamma = 2$, the
dimension of the substrate, we recover $\beta_{\rm rand}=1/2$ (larger
values of $\gamma$ result in the same $\beta$ since when $\gamma>2$
correlations become irrelevant). If $\gamma =
0$, the standard deviation of the populations growth rates has no
dependence on the population size ($\beta = 0$), as stated by Gibrat's
law, stating that the standard deviation does not depend in the cluster size. In the case of GB, $\gamma = 0.93\pm
0.08$ gives $\beta = 0.23\pm 0.02$ approximately consistent with the
measured value $\beta_{\rm GB}=0.27\pm 0.04$, within the error bars.
This observation suggests that the underlying demographic process
leading to the scaling in the standard deviation can be modeled as
arising from the long-range correlated growth of population cells.

\section{Discussion}

Our results suggest the existence of scale-invariant growth mechanisms
acting at different geographical scales. Furthermore,
Eq.~(\ref{scaling}) is similar to what is found for the growth of
firms and other macroeconomic indicators \cite{Stanley96,rossi1}.
Thus, our results support the existence of an underlying
link between the fluctuation dynamics of population growth and various
economic indicators, implying considerable unevenness in economic
development in different population sizes.
City growth is driven by many processes of which population growth and
migration is only one.  Our study captures only the growth of population
but not economic growth per se. Many cities grow economically while losing 
population and thus the processes we imply are those that influence a
changing population.  Our assumption is that population
change is an indicator of city growth or decline and therefore we have
based our studies on population clustering techniques. Alternatively,
the MSAs provides a set of rules that try to capture the idea of city
as a functional economic region.

The results we obtain
show scale-invariant properties which we have modelled using
long-range spatial correlations between the population of cells. That
is, strong development in an area attracts more development in its
neighborhood and much beyond. A key finding is that small places
exhibit larger fluctuations than large places.
The implications for locating activity in different places are that
there is a greater probability of larger growth in small places, but
also a greater probability of larger decline.
Opportunity must be weighed against the risk of failure.

One may take these ideas to a higher level of abstraction to study cell-to-cell flows (migration, commuting, etc.) gridded at different levels. As a consequence one may define population clusters, or MSAs, in terms of functional linkages between neighboring cells.
In addition one may relax some conditions imposed in the CCA. Here we consider a cell to be part of a cluster only if its population is strictly greater than 0. In SI Section V we relax this condition and study the robustness of the CCA when cells of a higher population than 0 (for instance, 5 and 20) are allowed into clusters and find that even though small clusters present a slight deviation, the overall behavior of the growth rate and standard deviation is conserved. 

\section{Materials and Methods}

\subsection{Information on the datasets}
\label{details}

The datasets analyzed in this paper were obtained from the websites
http://census.ac.uk, http://www.esri.com/, and
http://na.unep.net/datasets/datalist.php, for GB, USA and Africa,
respectively, and can be downloaded from
http://lev.ccny.cuny.edu/$\sim$hmakse/cities/city\_data.zip.

The datasets consist of a list of populations at specific coordinates
at two time steps $t_0$ and $t_1$. A graphical representation of the
data can be seen in Fig.~\ref{fig1}C for GB where each point represents 
a data point directly extracted from the dataset.

To perform the CCA at different scales we coarse-grain the
datasets. For this purpose, we overlay a grid on the corresponding map
(USA, GB, or Africa) with the desired cell size (for example, 2km-by-2km or 4km-by-4km for the USA).  Then, the population of each cell is
calculated as the sum of the populations of points (obtained from the
original dataset) that fall into this cell.

Table~\ref{table1} shows information on the datasets and results on
USA, GB and Africa for the cell size used in the main text as well as
some of the exponents obtained in our analysis.

\begin{table}[h]
\caption{\label{table1}
{\bf Characteristics of datasets and summary of results}}
\vspace{.5cm}
\begin{tabular}{|c|c|c|c|c|c|c|c|c| }
  \hline
  Data&Number &$t_0$ & $t_1$ &Average &Cell Size& Number of& $\alpha$ & $\beta$\\
  &of cells&& &growth rate& &  clusters& & \\
  \hline
  USA&   1.86 mill & 1990 & 2000 & 0.9\% & 2km-by-2km &  30,210 & 0.28 $\pm$ 0.08 & 0.20 $\pm$ 0.06\\
  \hline
  GB &   0.10 mill & 1981&1991& 0.3\% & 2.2km-by-2.2km &  10,178 &0.17 $\pm$ 0.05 & 0.27 $\pm$ 0.04\\
  \hline
  Africa &  2,216 & 1960&1990 &  4\% & 77km-by-77km & 3,988 & 0.21 $\pm$ 0.05 & 0.19 $\pm$ 0.04\\
  \hline
\end{tabular}
\end{table}

\subsection{Calculation of $\langle r(S_0)\rangle$ and $\sigma(S_0)$ and methodology}
\label{expla}

The average growth rate, $\langle r(S_0) \rangle =\ln(S_{1}/S_{0})$,
and the standard deviation, $\sigma(S_0) =\sqrt{\langle r(S_{0})^2
  \rangle - \langle r(S_{0}) \rangle^{2}}$, are defined as follows.  If we call $P(r | S_0)$ the
conditional probability distribution of finding a cluster with growth
rate $r(S_0)$ with the condition of initial population $S_0$, then we
can obtain $r(S_0)$ and $\sigma(S_0)$ through,
\begin{equation}
\langle r(S_{0}) \rangle = \int r P(r | S_0) {\rm d}r,
\end{equation}
and
\begin{equation}
\langle r(S_{0})^{2} \rangle = \int r^{2} P(r | S_0)
 {\rm d}r.
\end{equation}

Once $r(S_0)$ and $\sigma(S_0)$ are calculated for each cluster, we
perform a nonparametric regression
analysis~\cite{hardle,Silverman86}, a technique broadly used in the literature of population dynamics. The idea is to provide an estimate
for the relationship between the growth rate and $S_0$ and between the
standard deviation and $S_0$. Following the methods explained in
Ref.~\cite{Silverman86} we apply the Nadaraya-Watson method to
calculate an estimate for the growth rate, $\hat{r}(S_0)$, with,
\begin{equation}
  \langle  \hat{r}(S_0) \rangle=\frac{\sum_{i=0}^{\rm all clusters}
    K_h(S_0 - S_{i}(t_0))r_{i}(S_0)}{\sum_{i=0}^{\rm all clusters} K_h(S_0 -
    S_{i}(t_0))},
\end{equation}
and an estimate for the standard deviation $\hat{\sigma}(S_0)$ with,
\begin{equation}
\hat{\sigma}(S_0)=\sqrt{\frac{\sum_{i=0}^{\rm all clusters} K_h(S_0 - S_{i}(t_0))(r_{i}(S_0)-\langle  \hat{r}(S_0) \rangle)^2}{\sum_{i=0}^{\rm all clusters} K_h(S_0 - S_{i}(t_0))}},
\end{equation}
where $S_{i}(t_0)$ is the population of cluster $i$ at time $t_0$ (as defined in the main text), $r_{i}(S_0)$ is the growth rate of cluster $i$ and $K_h(S_0 - S_{i}(t_0))$ is a gaussian kernel of the form,
\begin{equation}
K_h(S_0 - S_{i}(t_0)) = {\rm e}^{\frac{({\rm ln}S_0 - {\rm ln}S_{i}(t_0))^2}{2h^2}}, \quad h=0.5
\end{equation}

Finally, we compute the 95\% confidence bands (calculated from 500
random samples with replacement) to estimate the amount of statistical
error in our results~\cite{hardle}. The bootstrapping technique was
applied by sampling as many data points as the number of clusters and
performing the nonparametric regression on the sampled data. By
performing 500 realizations of the bootstrapping algorithm and
extracting the so called $\alpha/2$ ($\alpha$ is not related to the growth rate exponent) quantile we obtain the 95\% confidence
bands.

To obtain the exponents $\alpha$ and $\beta$ of the power-law scalings
for $\langle r(S_0) \rangle$ and $\sigma(S_0)$, respectively, we
perform an OLS regression analysis~\cite{montgomery}. More
specifically, to obtain the exponent $\beta$ from Eq.~(\ref{scaling}),
we first linearize the data by considering the logarithm of the
independent and dependent variables so that Eq.~(\ref{scaling}) becomes
${\rm ln}~\sigma(S_0)\sim \beta~{\rm ln}~S_0$. Then, we apply a linear
Ordinary Least Square regression that leads to the exponent
\begin{equation}
  \beta = \frac{N_c \sum_{i=1}^{N_c} [{\rm ln}~S_{i}(t_0)~{\rm ln}~\sigma(S_{i}(t_0))] - \sum_{i=1}^{N_c} {\rm ln}~S_{i}(t_0) \sum_{i=1}^{N_c} {\rm ln}~\sigma(S_{i}(t_0))}{N_c \sum_{i=1}^{N_c} ({\rm ln}~S_{i}(t_0))^2 - (\sum_{i=1}^{N_c} {\rm ln}~S_{i}(t_0))^2},
\end{equation}
where $N_c$ is the number of clusters found using the CCA.  Analogously, we obtain the exponent $\alpha$ by
linearizing $\langle |r(S_0)| \rangle$ and calculating
\begin{equation}
  \alpha = \frac{N_c \sum_{i=1}^{N_c} ({\rm ln}~S_{i}(t_0)~{\rm ln}~\langle |r(S_{i}(t_0))| \rangle - \sum_{i=1}^{N_c} {\rm ln}~S_{i}(t_0) \sum_{i=1}^{N_c} {\rm ln}~\langle |r(S_{i}(t_0))| \rangle}{N_c \sum_{i=1}^{N_c} ({\rm ln}~S_{i}(t_0))^2 - (\sum_{i=1}^{N_c} {\rm ln}~S_{i}(t_0))^2}.
\end{equation}

Next we compute the 95\% confidence interval for the exponents
$\alpha$ and $\beta$. For this we follow the book of Montgomery and
Peck~\cite{montgomery}. The 95\% confidence interval for $\beta$ is
given by,
\begin{equation}
t_{0.025,N_c -2} * se,
\end{equation}
where $t_{\alpha' /2,N_c -2}$ is the t-distribution with parameters
$\alpha' /2$ and $N_c -2$ and $se$ is the standard error of the exponent
defined as
\begin{equation}
  se = \sqrt{\frac{SS_E}{(N_c-2)S_{xx}}},
\end{equation}
where $SS_E$ is the residual and $S_{xx}$ is the variance of $S_0$.

Finally, we express the value of the exponent in terms of the 95\% confidence intervals as,
\begin{equation}
\beta \pm t_{0.025,N_c -2} * se.
\end{equation}

\subsubsection*{Acknowledgments}

We thank 
L.H. Dobkins and J. Eeckhout for providing data on MSA and C. Briscoe
for help with the manuscript.  This work is supported by the National
Science Foundation through grant NSF-HSD.  J.S.A. thanks the Brazilian
agencies CNPq, CAPES, FUNCAP and FINEP for financial support.

\newpage

\begin{figure}[h]
\caption{{\bf (A)} Sketch illustrating the CCA 
applied to a sample of gridded population data. In the top left panel,
cells are colored in blue if they are populated ($n_j^{(i)}(t) > 0$),
otherwise, if $n_j^{(i)}(t)=0$, they are in white. In the top right panel we
initialize the CCA by selecting a populated cell and
burning it (red cell). Then, we burn the populated neighbors of the
red cell as shown in the lower left panel. We keep growing the cluster
by iteratively burning neighbors of the red cells until all
neighboring cells are unpopulated, as shown in the lower right
panel. Next, we pick another unburned populated cell and repeat the
algorithm until all populated cells are assigned to a cluster.
The population $S_i(t)$ of cluster $i$ at time $t$ is then
$S_i(t)=\sum_{j=1}^{N_i} n_j^{(i)}(t)$.  {\bf (B)} Cluster identified with
the CCA in the London area (red) overlaying a
corresponding satellite image (extracted from maps.google.com).  The
greenery corresponds to vegetation, and thus approximately indicates
unoccupied areas.  For example, Richmond Park can be found as a
vegetation area in the south-west.  The areas in the east along the
Thames River correspond mainly to industrial districts and in the west
the London Heathrow Airport, also not populated.  The yellow line in
the center represents the administrative boundary of the City of London,
demonstrating the difference with the urban cluster found
with the CCA.  The pink clusters surrounding the major
red cluster are smaller conglomerates not connected to London. The
figure shows that an analysis based on the City of London captures
only a partial area of the real urban agglomeration.
{\bf (C)} Result of the CCA applied to all of GB showing the
large variability in the population distribution. The color bar (in
logarithmic scale) indicates the population of each urban cluster.
}
\label{fig1}
\end{figure}

\begin{figure}[h]
\caption{ Illustration of possible changes in cluster shapes. 
In each case we show how the growth rate is computed. 
In the first case, there is no areal modification in
the cluster between $t_0$ and $t_1$. In the second, the cluster
expands. In the third the cluster reduces its area. In the fourth, one
cluster divides into two and therefore we consider the population at
$t_1$ to be $S_1 = S'_1+S''_1$. In the fifth case two clusters merge
to form one at $t_1$. In this case we consider the population at $t_0$
to be $S_0 = S'_0+S''_0$.
}
\label{cases}
\end{figure}

\begin{figure}
\caption{Results for the USA using a cell size of 2000m-by-2000m. {\bf (A)}
Mean annual growth rate for population clusters in the USA versus
initial population of the clusters. The straight dashed line shows a
power-law fit 
with $\alpha_{\rm USA} =0.28\pm0.08$ as determined using OLS
regression. {\bf (B)} Standard deviation of the growth rate for the
USA. The straight dashed line corresponds to a power-law fit using OLS
regression with $\beta_{\rm USA}=0.20\pm0.06$.
}
\label{usascaling}
\end{figure}

\begin{figure}
\caption{Results for Great Britain using a cell size of 2.2km-by-2.2km. {\bf
  (A)} Mean annual growth rate of population clusters in Great Britain
versus the initial cluster population. The inset shows a double
logarithmic plot of the growth rate in the intermediate range of
populations, $3000<S_0<3 \times 10^5$.  A power-law fit using OLS
leads to an exponent $\alpha_{\rm GB}=0.17\pm0.05$ for this
range. {\bf (B)} Double logarithmic plot of the standard deviation of
the annual growth rates of population clusters in Great Britain versus
the initial cluster population. The straight line corresponds to a
power-law fit using OLS with an exponent $\beta_{\rm GB}=0.27\pm0.04$,
according to Eq.~(\ref{scaling}).
{\bf (C)} Scaling of the standard deviation in cluster population
obtained from the randomized surrogate dataset of GB by randomly
swapping the cells. The data shows an exponent $\beta_{\rm rand}=1/2$
in the tail. The deviations for small $S_0$ are discussed in the SI
Section~IV. where we test these results by generating random
populations. {\bf (D)} Long-range spatial correlations in the
population growth of cells for GB according to Eq. (\ref{cell}). The
straight line corresponds to an exponent $\gamma=0.93\pm 0.08$.
}
\label{ukscaling}
\end{figure}

\begin{figure}
\caption{ Results for Africa using a cell size of 77km-by-77km. {\bf (A)}
Mean growth rate of clusters in Africa versus the initial size of
population $S_0$. The straight dashed line shows a power-law fit with
exponent $ \alpha_{\rm Af} = 0.21 \pm0.05$, obtained using OLS
regression.  {\bf (B)} Standard deviation of the growth rate in
Africa. The straight line corresponds to power-law fit using OLS
providing the exponent $\beta_{\rm Af}=0.19\pm0.04$.
}
\label{africascaling}
\end{figure}

\setcounter{figure}{0}
\begin{figure}[h]
{\bf A} \includegraphics[width=0.7\textwidth]{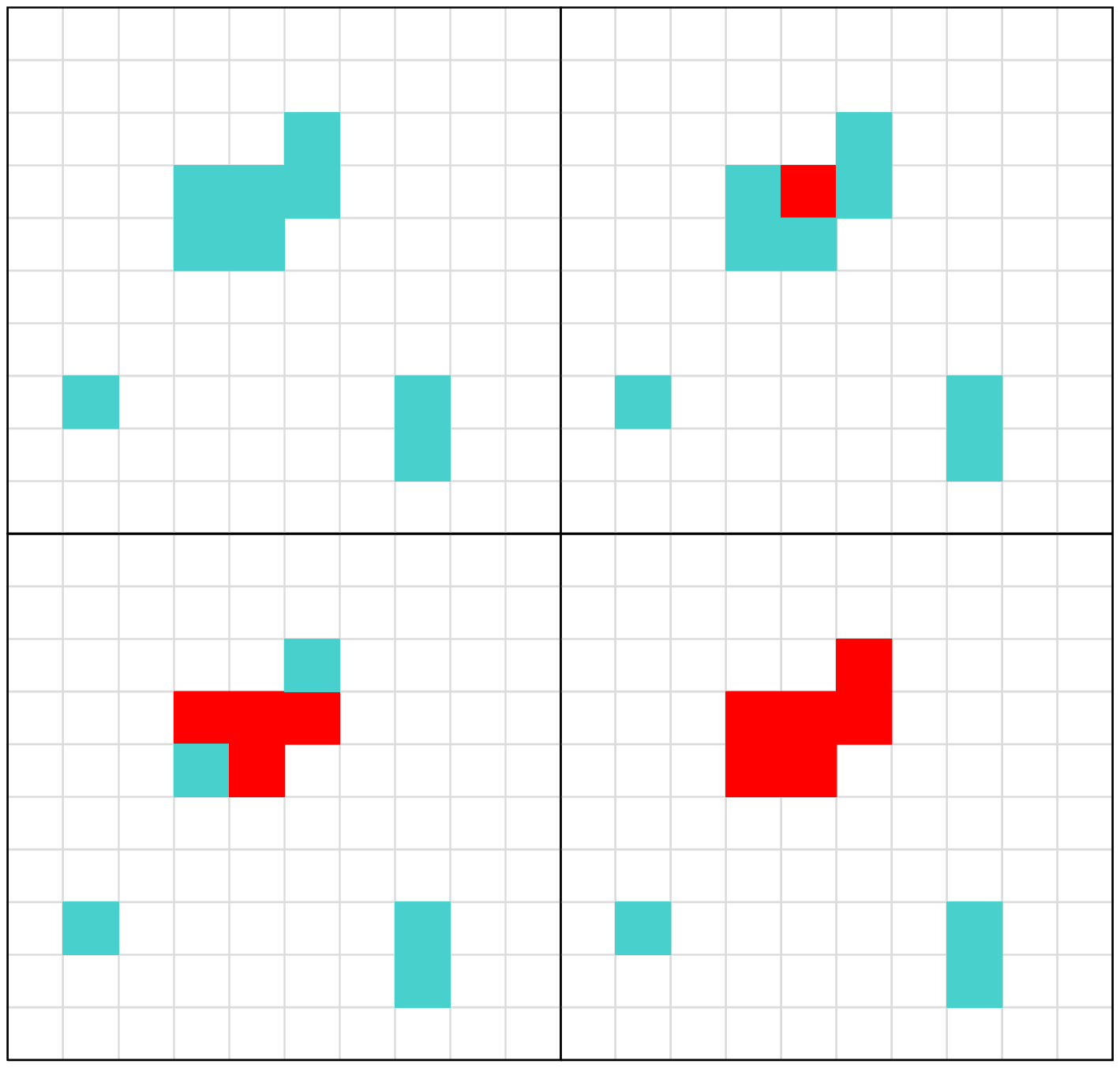}
\renewcommand{\figurename}{{\bf Fig.}}
\caption{}
\label{fig1}
\end{figure}

\newpage

\setcounter{figure}{0}
\begin{figure}[h]
{\bf B} \includegraphics[width=0.5\textwidth]{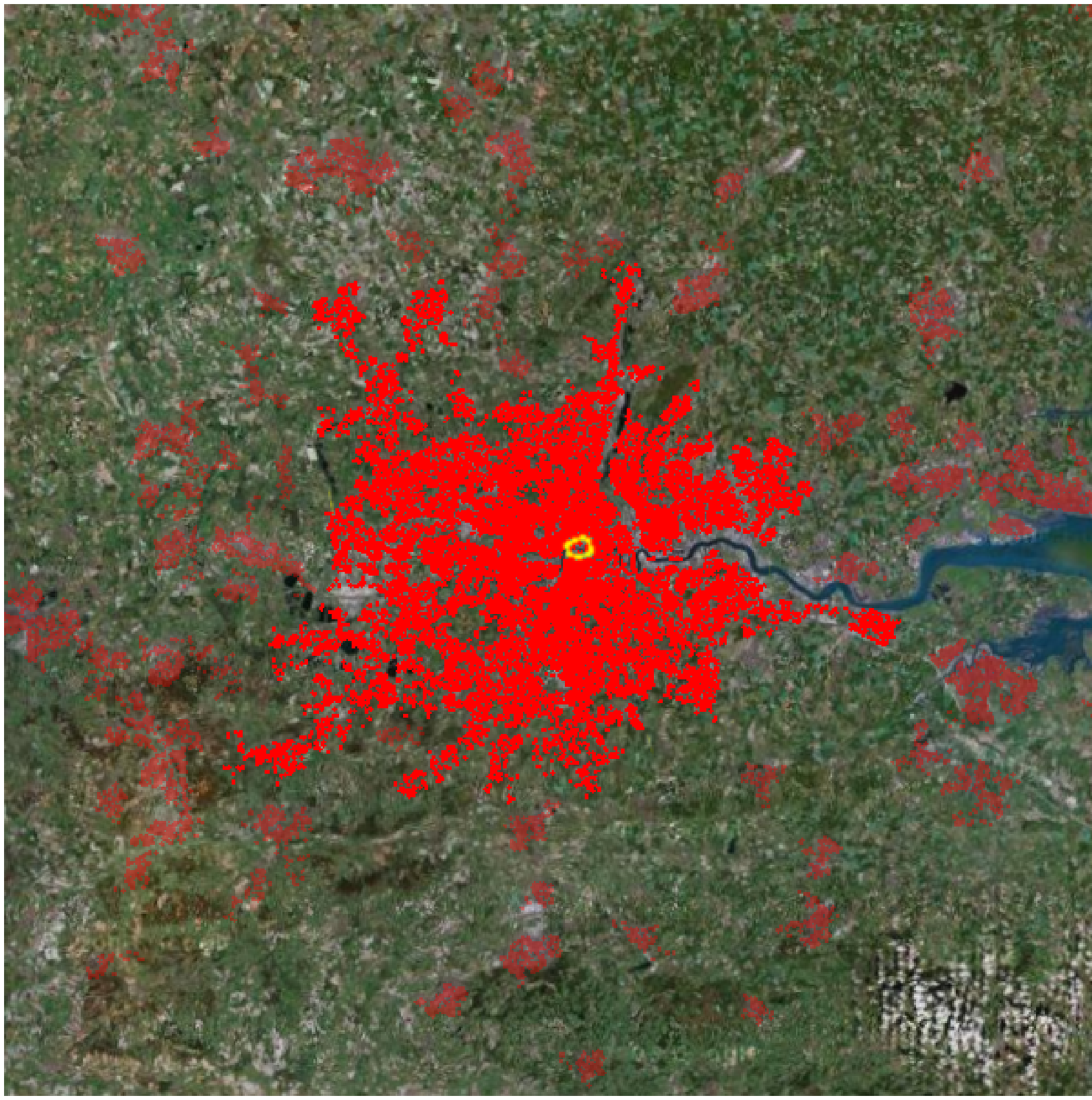}

{\bf Fig.~\ref{fig1}}
\end{figure}

\newpage

\setcounter{figure}{0}
\begin{figure}[h]
{\bf C} \includegraphics[width=0.6\textwidth]{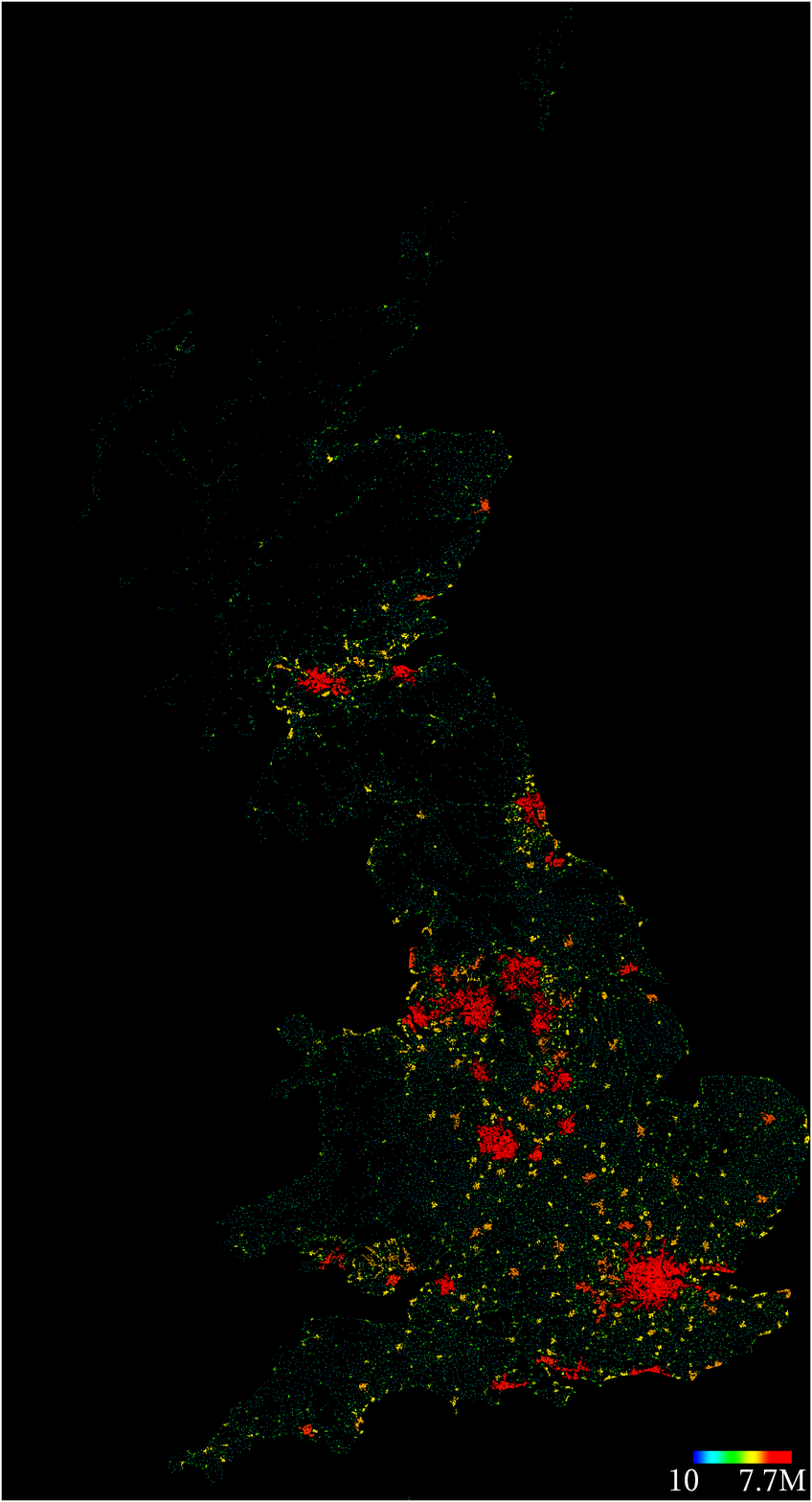}

{\bf Fig.~\ref{fig1}}
\end{figure}

\newpage

\setcounter{figure}{1}
\begin{figure}[h]
\includegraphics[width=0.7\textwidth]{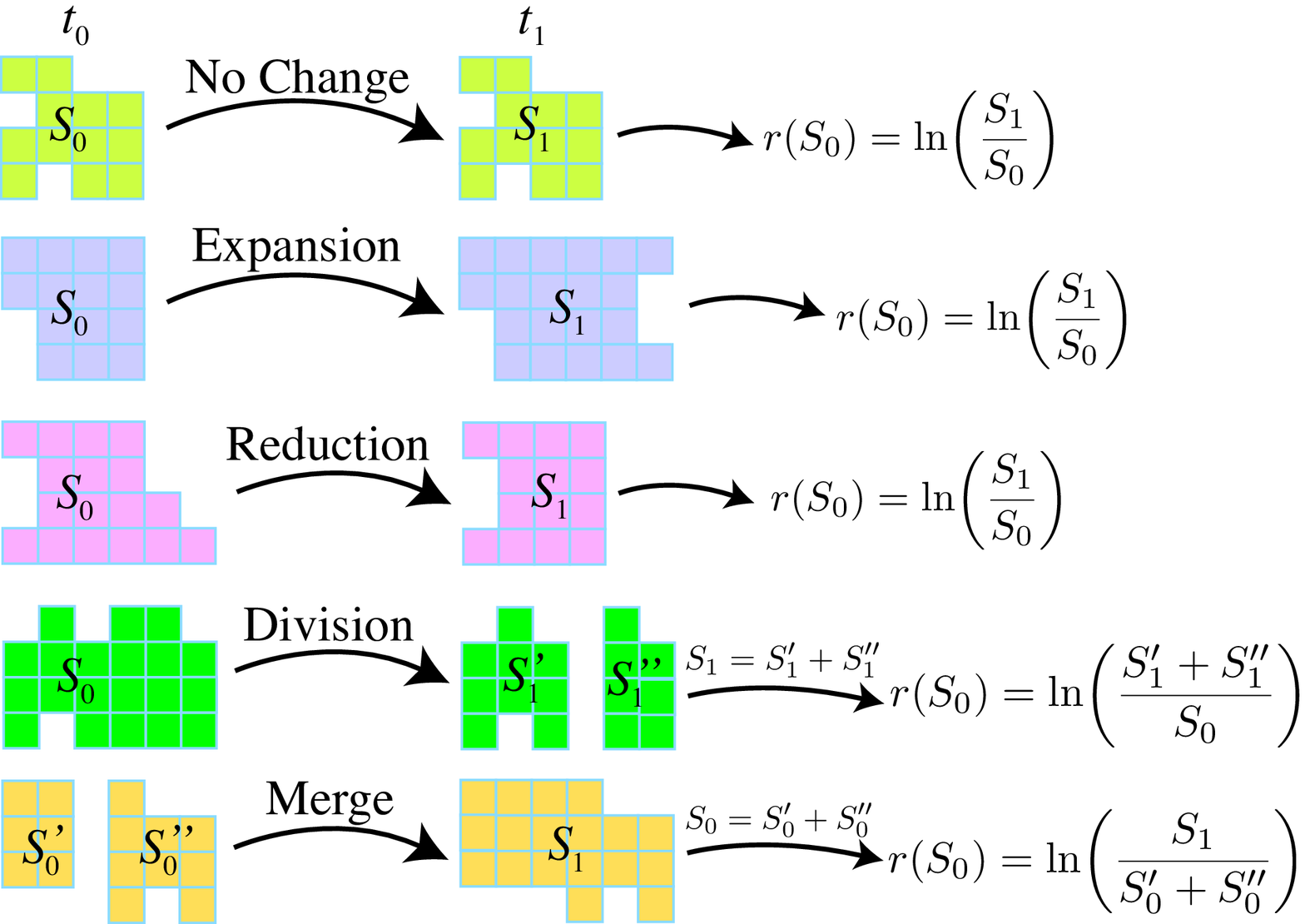}
\renewcommand{\figurename}{{\bf Fig.}}
\caption{}
\label{cases}
\end{figure}

\clearpage
\newpage

\setcounter{figure}{2}

\begin{figure}
\centering {
\hbox{{\bf A}
\resizebox{0.45\textwidth}{!}{\includegraphics{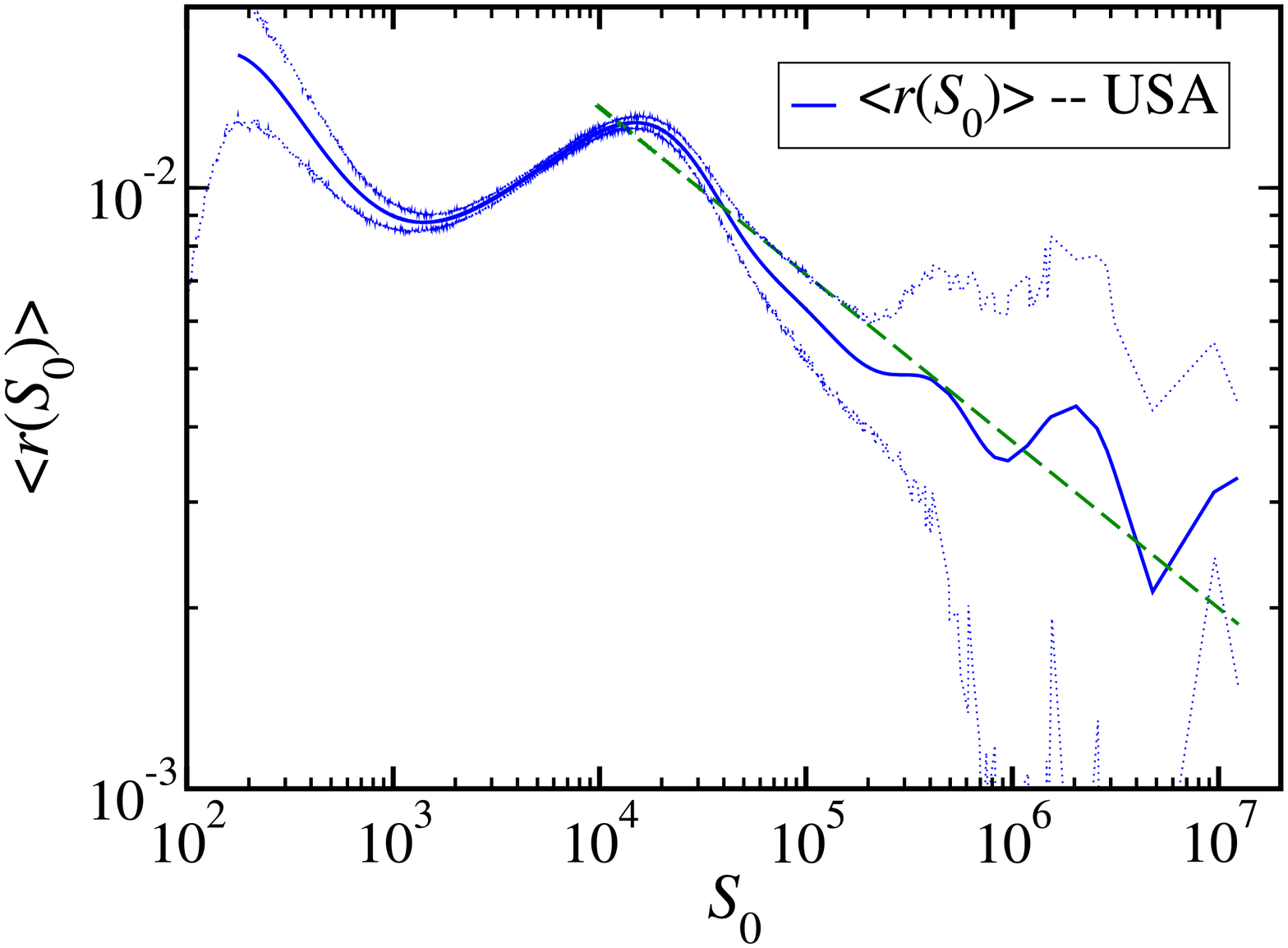}}
\hspace{1cm}
{\bf B} \resizebox{0.45\textwidth}{!}{\includegraphics{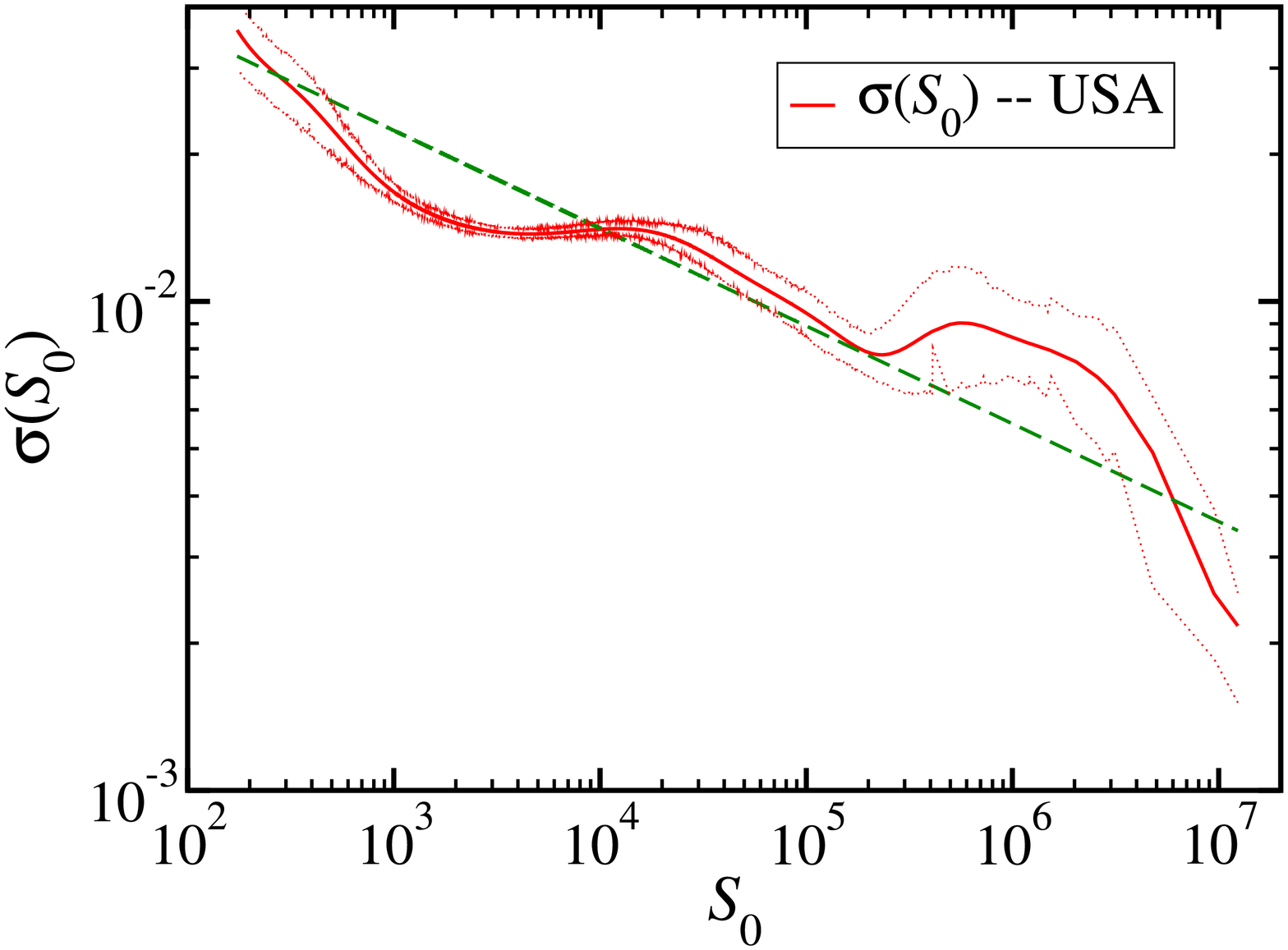}}
}
}
\caption{}
\label{usascaling}
\end{figure}

\clearpage
\newpage

\begin{figure}
\centering {
\hbox{{\bf A}
\resizebox{0.45\textwidth}{!}{\includegraphics{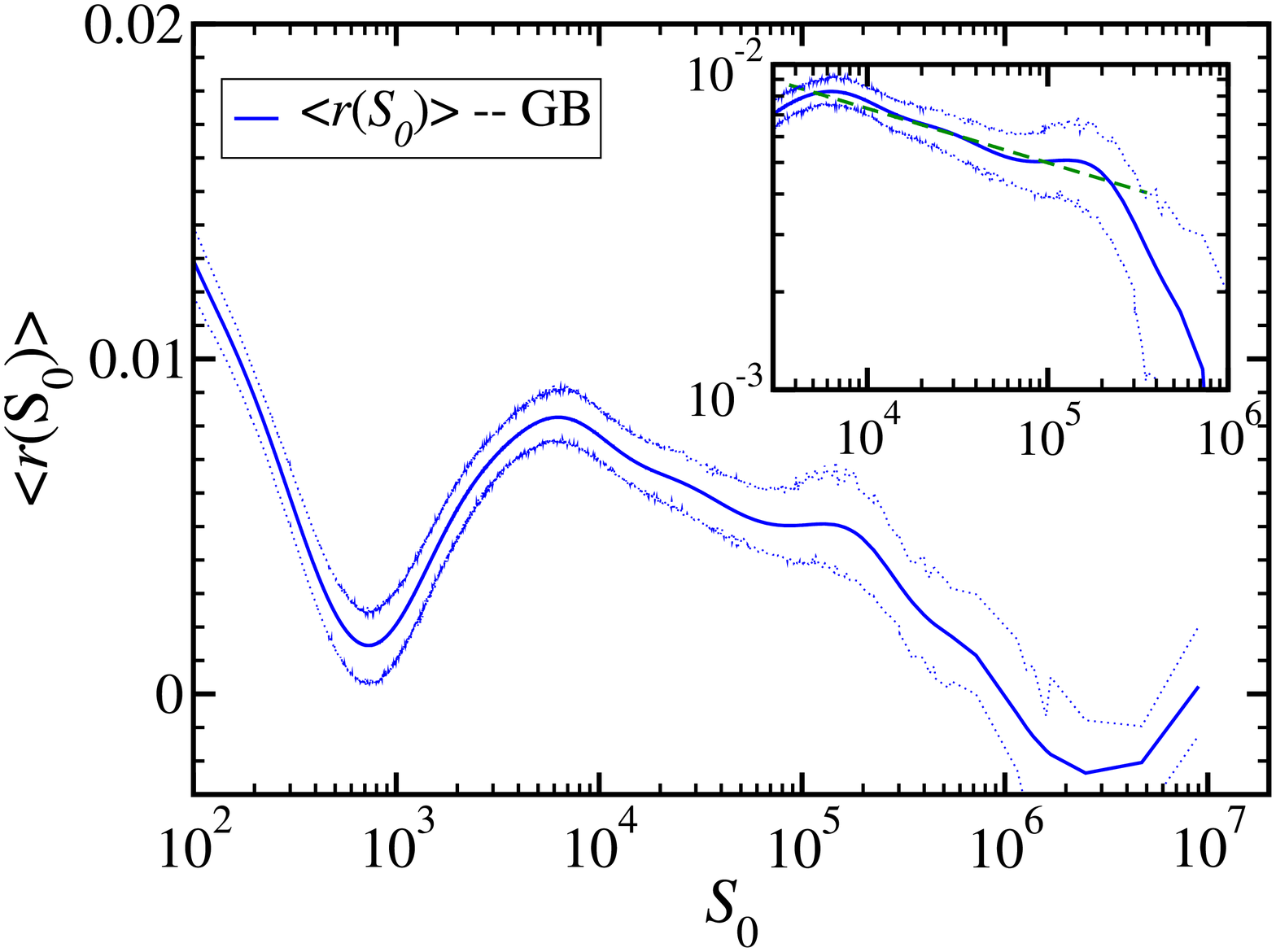}}
\hspace{1cm}
{\bf B} \resizebox{0.45\textwidth}{!}{\includegraphics{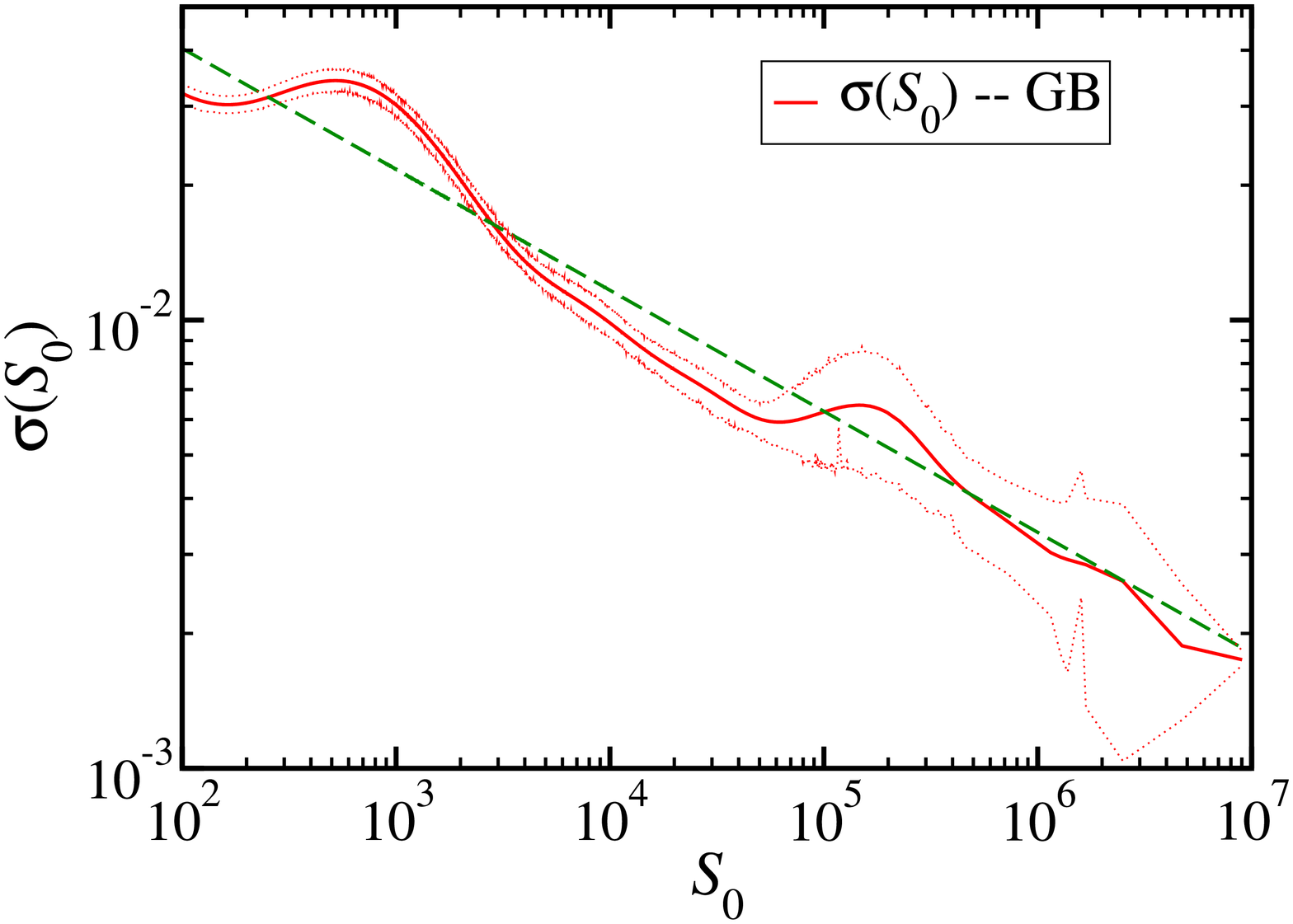}}
}
\vspace{1cm}
\hbox{
{\bf C} \resizebox{0.45\textwidth}{!}{\includegraphics{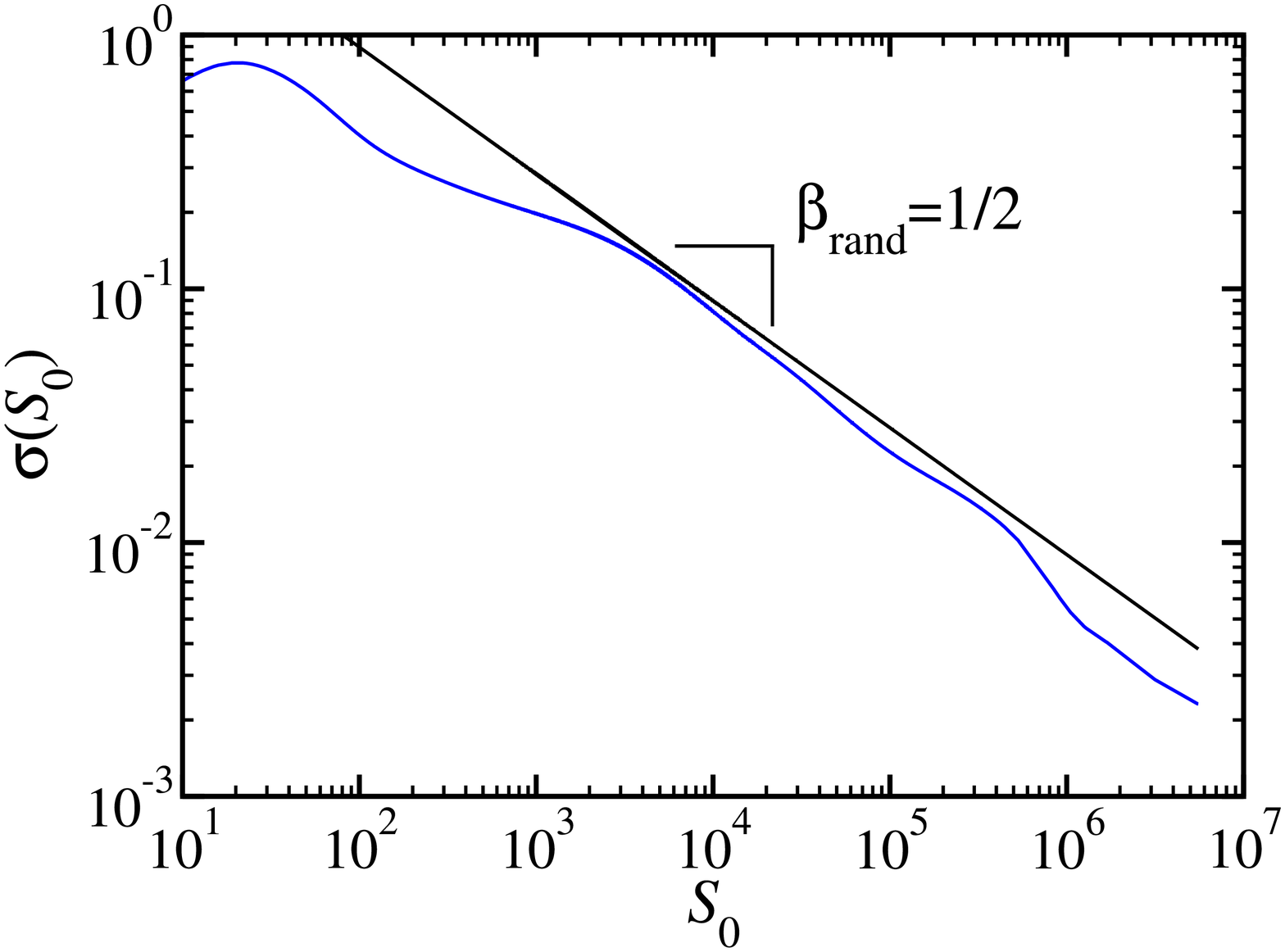}}
\hspace{1cm}
{\bf D} \resizebox{0.45\textwidth}{!}{\includegraphics{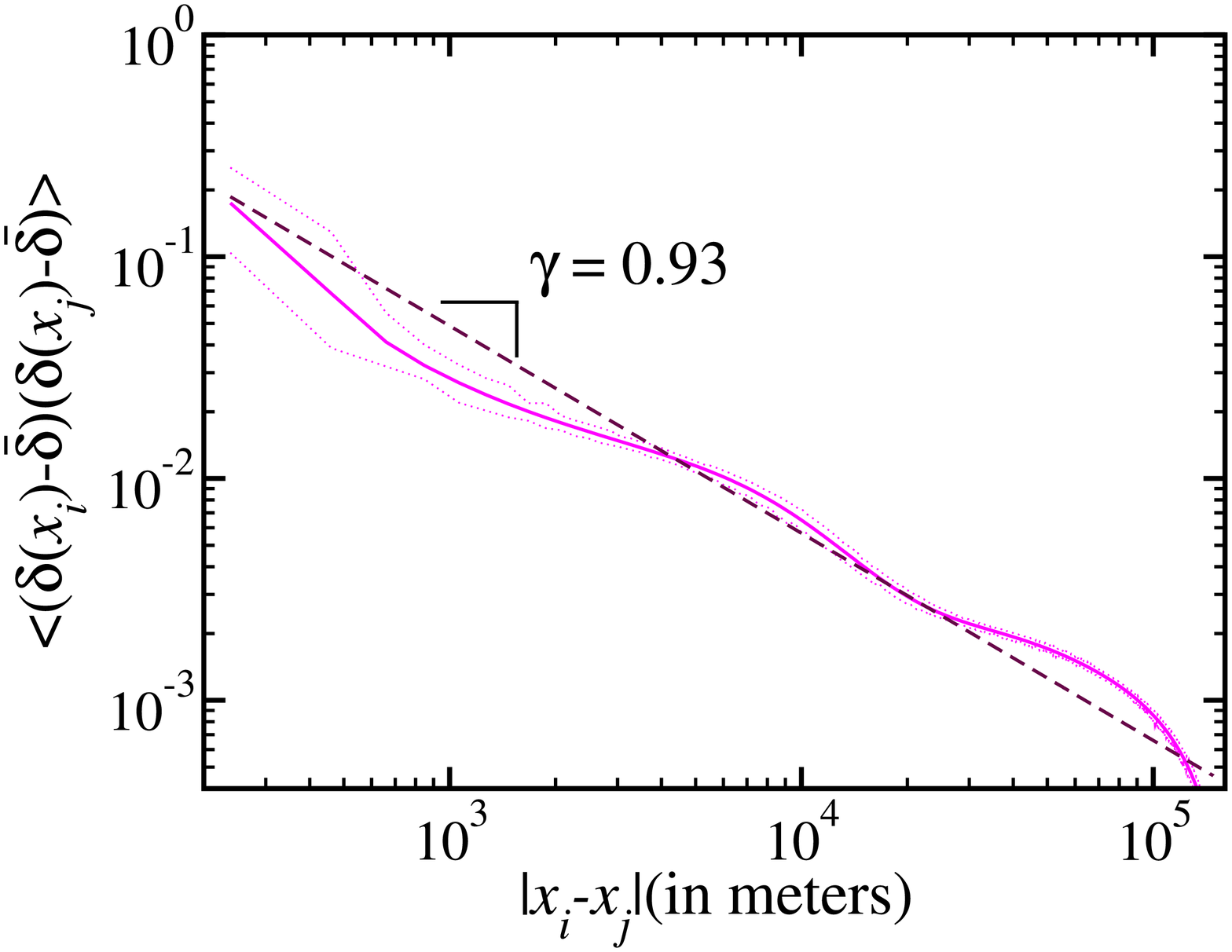}}
}
}
\caption{}
\label{ukscaling}
\end{figure}

\clearpage
\newpage

\begin{figure}
\centering {
\hbox{{\bf A}
\resizebox{0.45\textwidth}{!}{\includegraphics{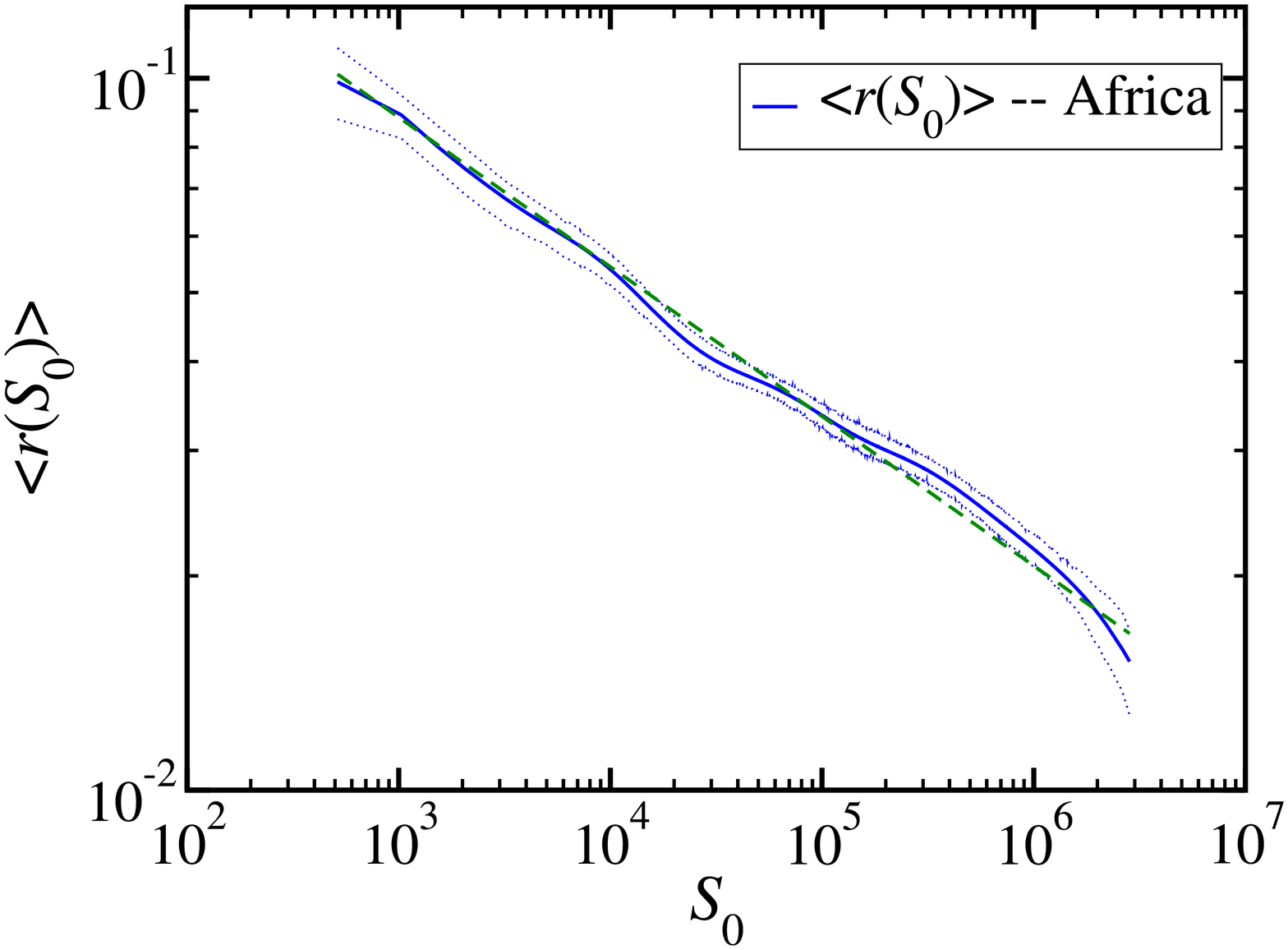}}
\hspace{1cm}
{\bf B} \resizebox{0.45\textwidth}{!}{\includegraphics{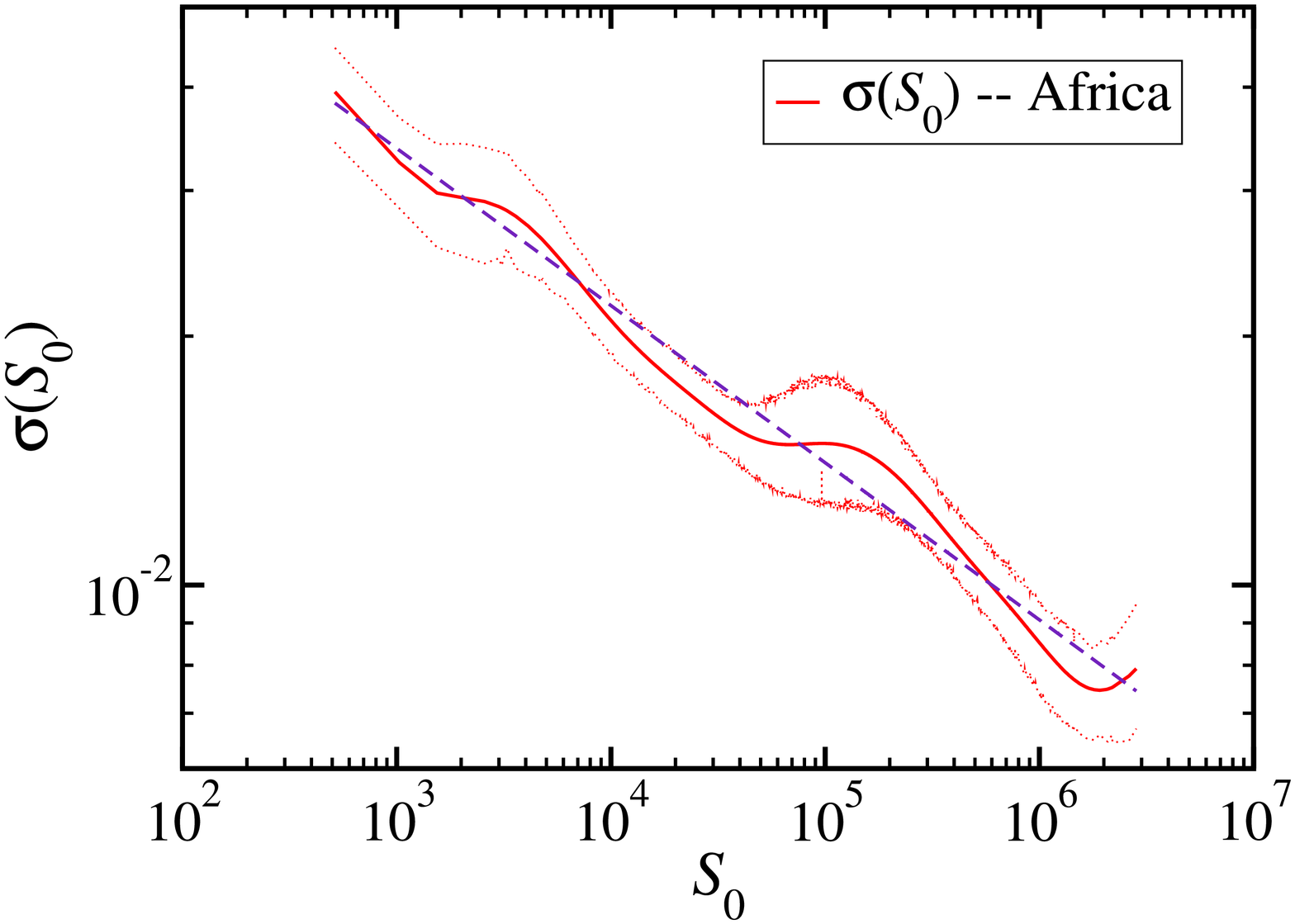}}
}
}
\caption{}
\label{africascaling}
\end{figure}

\clearpage
\newpage

\centerline{\bf SUPPORTING INFORMATION}

\vspace{0.5cm}

\begin{center}{\bf  Laws of Population Growth}

\vspace{0.5cm}

Hern\'an D. Rozenfeld, Diego Rybski,
Jos\'e S. Andrade Jr., \\Michael Batty,
H. Eugene Stanley, and Hern\'an A. Makse
\end{center}

\vspace{0.5cm}

As supplementary materials we provide the following:
In Section~\ref{comparison} we present tables with details on our
results using the CCA and results presented in previous papers to
allow for comparison between the different approaches.
In Section~\ref{robustness} we study the stability of the scaling
found in the text under a change of scale in the cell size.  In
Section~\ref{sec:si:correlations} we detail the calculations to relate
spatial correlations between the population growth and $\sigma(S_0)$
namely the relation $\beta=\gamma /4$.  In Section~\ref{random} we
describe the random surrogate dataset used to further test our
results. In Section~\ref{variation} we further test the robustness of the CCA by proposing a small variation in the algorithm.

\section{Clusters at different scales and comparison with Metropolitan
  Statistical Areas}
\label{comparison}

In this section, Tables S1 and S2 allow for a
detailed comparison of urban clusters obtained with the CCA applied to
the USA in 1990, and the populations of MSA from US Census Bureau used
in previous studies of population growth
~\cite{dobkins00,Ioa03,Eeckhout04}.

We can see that the MSA presented by Eeckhout (2004) typically
correspond to our clusters using cell sizes of 4km and 8km. For example,
for the New York City region Eeckhout's data are well approximated by a
cell size of 4km, but Los Angeles is better approximated when using a
cell size of 8km. On the other hand Dobkins-Ioannides (2000) data are
better described by cell sizes of 2km or 4km. For instance, Chicago is
well described by a cell size of 4km and Los Angeles is better
described by a cell size of 2km.

An interesting remark is that the population of Los Angeles when using
cell sizes of 2km, 4km and 8km does not vary as much as that for New
York. This could be caused by the fact that major cities in the
northeast of USA are closer to each other than large cities in the
southwest, which may be attributed to land or geographical
constraints.

It is important relate the results of Table S2 with an ecological fallacy. As the cell size is increased, the population of a cluster also increases, as expected, because the cluster now covers a larger area. This is not a direct manifestation of an ecological fallacy which, would appear if the statistical results (growth rate vs. S
or standard deviation vs. S) gave different results as the cell size
increases. In Fig. 1 and Fig. 2 in the SI Section~\ref{robustness}, we observe that the growth rate and
standard deviation for the USA and GB follow the same form, except for the
case of the growth rate in the USA in which different cell sizes show
deviations from each other. The later may be an indicative of an ecological
fallacy. In this case, it is not obvious what cell size is the ``correct'' one.
We consider this point (the possibility to choose the cell size) to be a feature of the CCA, since one may appropriately
pick the cell size according to the specific problem one is studying.

Table S1: {\bf Top 10 largest MSA of the USA in 1990 from previous analysis of population growth}

\begin{table}[h]
   \vspace{.5cm}
  \begin{tabular}{|c|c|c|c|c|}
    \hline
    &\multicolumn{2}{|c|}{Dobkins - Ioannides}&\multicolumn{2}{|c|}{Eeckhout}\\
    \hline
    &MSA&Population& MSA&Population\\
    \hline
    1&NYC NY206& 9,372,000&NYC-North NJ-Long Is., NY-NJ-CT-PA & 19,549,649\\
    \hline
    2&Los Angeles CA172&8,863,000&Los Angeles-Riverside-Orange County, CA &14,531,529\\
    \hline
    3&Chicago IL59&7,333,000&Chicago-Gary-Kenosha, IL-IN-WI&8,239,820\\
    \hline
    4&Philadelphia PA228&4,857,000&Washington-Baltimore, DC-MD-VA-WV&6,727,050\\
    \hline
    5&Detroit MI80&4,382,000&San Francisco-Oakland-San Jose, CA&6,253,311\\
    \hline
    6&Washington DC312&3,924,000&Philadelphia-Wilmington-Atlantic City&5,892,937\\[-2ex]
    &&&PA-NJ-DE-MD&\\
    \hline
    7&San Francisco CA266&3,687,000&Boston-Worcester-Lawrence, MA-NH-ME-CT&5,455,403\\
    \hline
    8&Houston TX129&3,494,000&Detroit-Ann Arbor-Flint, MI&5,187,171\\
    \hline
    9&Atlanta GA19&2,834,000&Dallas-Fort Worth, TX&4,037,282\\
    \hline
    10&Boston MA39&2,800,000&Houston-Galveston-Brazoria, TX&3,731,131\\
    \hline
\end{tabular}
\end{table}

\newpage

Table S2: {\bf Top 10 largest clusters of the USA in 1990 from our analysis for different cell sizes.} The city names are the major cities that belong to the clusters and were picked to show the areal extension of the cluster.

\begin{table}[h]
\vspace{.5cm}
\begin{tabular}{|c|c|c|c|c|c|c|c|c|}
\hline
 &  \multicolumn{2}{|c|}{Cell = 1km}&\multicolumn{2}{|c|}{Cell = 2km}&\multicolumn{2}{|c|}{Cell = 4km}&\multicolumn{2}{|c|}{Cell = 8km}\\
\hline
&Cluster&Population& Cluster&Population&Cluster&Population& Cluster&Population\\
\hline
1&NYC & 7,012,989&NYC-Long Is.&12,511,237&NYC-Long Is.&17,064,816&NYC-Long Is.&41,817,858\\[-2ex]
& & &Newark&& N. NJ-Newark&&North NJ&\\[-2ex]
& & & Jersey City&& Jersey City&&Philadelphia&\\[-2ex]
& & & && &&D.C.-Boston&\\
\hline
2&Chicago&2,312,783&Los Angeles&9,582,507&Los Angeles&10,878,034&Los Angeles&13,304,233\\[-2ex]
&&&Long Beach&&Long Beach&&San Clemente&\\[-2ex]
&&&&&Pomona&&Riverside&\\
\hline
3&Los Angeles&1,411,791&Chicago&4,836,529&Chicago&7,230,404&Chicago&9,288,345\\[-2ex]
&&&Rockford&&Gary&&Gary&\\[-2ex]
&&&&&Rockford&& Rockford&\\[-2ex]
&&&&&&&Milwaukee&\\
\hline
4&Philadelphia&1,282,834&Philadelphia&3,151,704&Washington&5,316,890&San Francisco&5,736,479\\[-2ex]
&&&Wilmington&&Baltimore&&Santa Cruz&\\[-2ex]
&&&&&Springfield&&Brentwood&\\
\hline
5&Boston&759,024&Detroit&2,906,453&Philadelphia&4,935,734&Detroit&4,442,723\\[-2ex]
&&&&&Trenton&&Ann Arbor&\\[-2ex]
&&&&&Wilmington&&Monroe&\\[-2ex]
&&&&&&&Sarnia&\\
\hline
6&Newark&581,048&San Francisco&2,601,639&San Francisco&4,766,960&Miami&4,000,432\\[-2ex]
&&&San Jose&&San Jose&&Port St. Lucie&\\[-2ex]
&&&&&Concord&&&\\
\hline
7&San Francisco&507,300&Washington&2,059,421&Detroit&3,722,778&Dallas&3,536,186\\[-2ex]
&&&Alexandria&&Waterford&&Fort Worth&\\[-2ex]
&&&Bethesda&&Canton&&&\\
\hline
8&Washington&504,068&Phoenix&1,556,077&Miami&3,719,773&Houston&3,425,647\\[-2ex]
&&&&&W. Palm Beach&&&\\
\hline
9&Jersey City&438,591&Boston&1,498,208&Dallas&3,134,233&Cleveland&3,233,341\\[-2ex]
&&&Lowell&&Fort Worth&&Canton&\\[-2ex]
&&&Quincy&&&&&\\
\hline
10&Baltimore&437,413&Miami&1,465,490&Boston&3,064,925&Pittsburgh&3,214,661\\[-2ex]
&&&&&Brockton&&Youngstown&\\[-2ex]
&&&&&Nashua&&Morgantown&\\
\hline
\end{tabular}
\end{table}

\section{Scaling under coarse-graining}
\label{robustness}

In this section we test the sensitivity of our results to a
coarse-graining of the data. We analyze the average growth rate
$\langle r(S_0) \rangle$ and the standard deviation $\sigma(S_0)$ for
GB and the USA by coarse-graining the data sets at different levels.

\begin{figure}
\centering {
\hbox{{\bf A} \resizebox{0.5\textwidth}{!}{\includegraphics{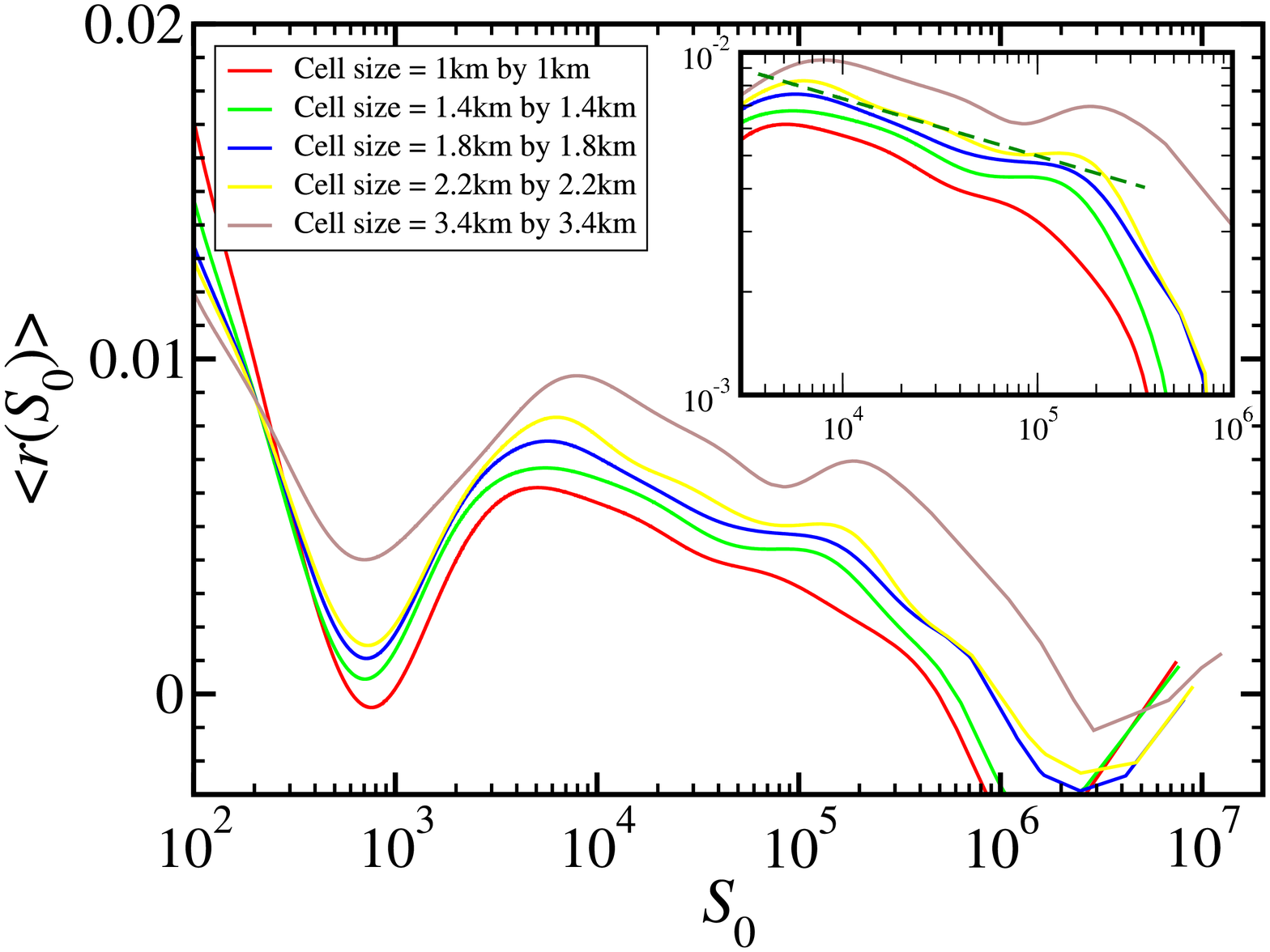}}
{\bf B} \resizebox{0.5\textwidth}{!}{\includegraphics{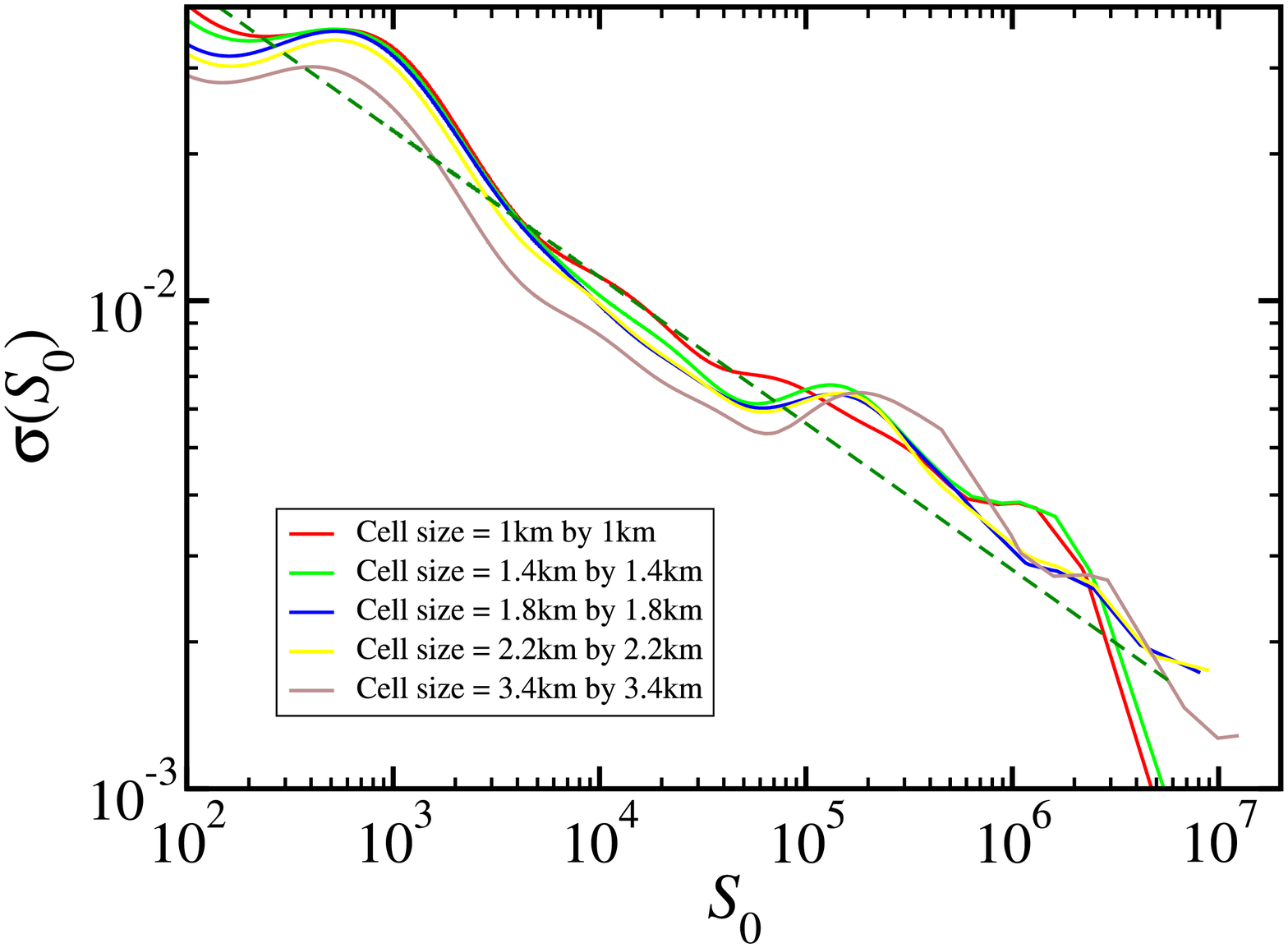}}
}}
\caption{Sensitivity of the results under coarse-graining of the
    data for GB. {\bf (A)} Average growth rate and {\bf (B)} standard deviation
  for GB using the clustering algorithm for different cell size. The
  dashed line represents the OLS regression estimate for the exponents
  {\bf (A)} $\alpha_{\rm GB}=0.17$ and {\bf (B)} $\beta_{\rm GB}=0.27$ obtained in
  the main text. For clarity we do not show the confidence bands. }
\label{uk_coarse}
\end{figure}

\begin{figure}
\centering {
\hbox{{\bf A} \resizebox{0.5\textwidth}{!}{\includegraphics{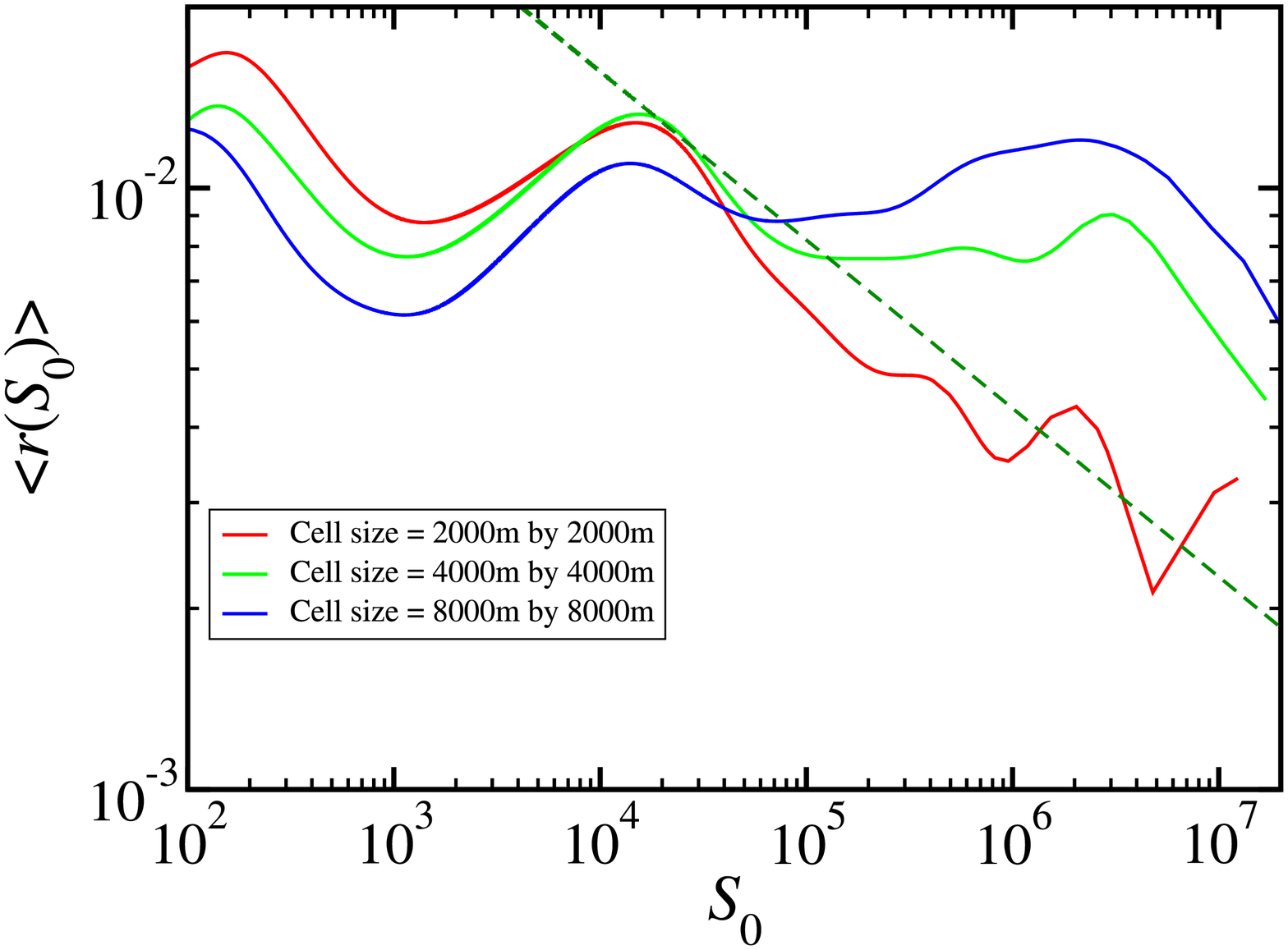}}
{\bf B} \resizebox{0.5\textwidth}{!}{\includegraphics{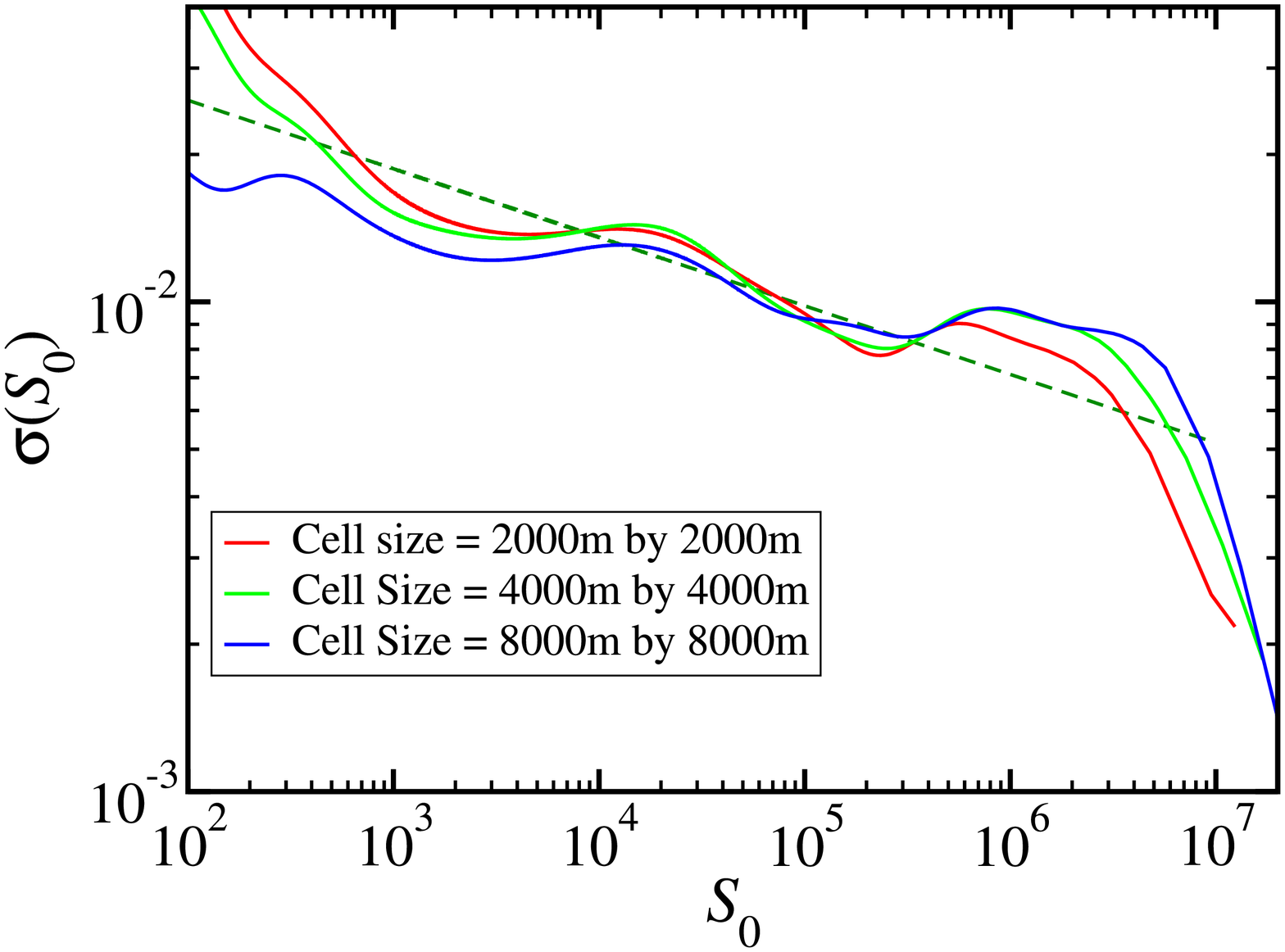}}
}}
\caption{Study of results under coarse-graining of the data for the USA.
{\bf (A)} Average growth rate and {\bf (B)} standard deviation for the USA
  using the clustering algorithm for different cell size. The dashed
  line represents the OLS regression estimate for the exponents {\bf (A)}
  $\alpha_{\rm USA}=0.28$ and {\bf (B)} $\beta_{\rm USA}=0.20$ obtained in
  the main text. For clarity we do not show the confidence bands.}
\label{usa_coarse}
\end{figure}

In Fig.~\ref{uk_coarse}A we observe that although the results are not
identical for all coarse-grainings, they are statistically similar,
showing a slight decay in the growth rate. Moreover, we see that
cities of size $S_0 \approx 10^3$ and $S_0 \approx 10^6$ still exhibit
a tendency to have negative growth rates for all levels of
coarse-graining, as explained in the main text. In the case of the USA
(Fig.~\ref{usa_coarse}A) there is a crossover to a flat behavior at a
cell size of 8000m, although at this scale all the northeast USA
becomes a large cluster of 41 million inhabitants. On the other hand,
Figs.~\ref{uk_coarse}B,~\ref{usa_coarse}B show that the scaling of
Eq.~(3) in the main text, $\sigma(S_0) \sim S_{0}^{-\beta}$, still holds when using the coarse-grained datasets on both GB and the USA.

\section{Correlations}
\label{sec:si:correlations}

In this section we elaborate on the calculations leading to the
relation between Gibrat's law and the spatial correlations in the cell
population.  We first show that when the population cells are randomly
shuffled (destroying any spatial correlations between the growth rates
of the cells), the standard deviation of the growth rate becomes
$\sigma(S_0) \sim S_0^{-\beta_{\rm rand}}$, where $\beta_{\rm rand} =
1/2$ \cite{Stanley96}.
Then, we show that long-range spatial
correlations in the population of the cells leads to the relation
$\beta=\gamma/4$ as stated at the end of Section II in the main text.

Assuming that the population growth rate is
small ($r \ll 1$), we can write $R = e^r \approx 1+r$. Replacing $R = 1+r$ in
Eq.~(1) in the main text we obtain
\begin{equation}
S_1 = S_0 + S_0 r.
\end{equation}

We define the standard deviation of the populations $S_1$ as $\sigma_1$,
which is a function of $S_0$:

\begin{equation}
\sigma_{1}(S_0) = \sqrt{\langle S_{1}^2 \rangle - \langle S_{1}
  \rangle^2}.
\end{equation}
This quantity is easier to relate to the spatial correlations of the
cells than the standard deviation $\sigma(S_0)$ of the growth rates
$r$.  Then, since $\langle S_{1} \rangle = S_{0} + S_{0} \langle r
\rangle$ and $\langle S_{1}^2 \rangle = S_{0}^2 + 2S_{0}^2 \langle r
\rangle + S_{0}^2 \langle r^2 \rangle$, we obtain,

\begin{equation}
\sigma_1(S_0) \sim
S_{0} \sigma(S_0),
\end{equation}
 where $\sigma(S_0) = \sqrt{\langle r^2 \rangle -
  \langle r \rangle^2}$ as defined in the main text.  Therefore, using
Eq.~(3) in the main text,
\begin{equation}
\label{eq:Sigma_exponent} \sigma_1(S_0) \sim S_{0}^{1-\beta}.
\end{equation}

As stated in the main text, the total population of a cluster at time $t_{0}$
is the sum of the populations of each cell, $S_{0} = \sum_{j=1}^{N_i} n_j^{(i)}$,
where $N_i$ is the number of cells in cluster $i$. The population of a cluster
at time $t_1$ can be written as
\begin{equation}
S_1 = S_0 + \sum_{j=1}^{N_i} \delta_j,
\end{equation}
where $\delta_j$ is the increment in the population of cell $j$ from
time $t_0$ to $t_1$ (notice that $\delta_j$ can be
negative). Therefore, the standard deviation $\sigma_1(S_0)$ is
\begin{equation}
  \label{eq:Sigma} \Big(\sigma_1(S_0)\Big)^2 = \sum_{j,k}^{N_i} \langle \delta_j \delta_k
  \rangle - \langle\sum_{j}^{N_i} \delta_j\rangle^{2} = \sum_{j,k}^{N_i} \langle (\delta_j -
  \bar{\delta}) (\delta_k - \bar{\delta}) \rangle.
\end{equation}

After the process of randomization explained in Section II main text, the
correlations between the increment of population in each cell are destroyed.
Thus,
\begin{equation}
\langle (\delta_j - \bar{\delta}) (\delta_k - \bar{\delta}) \rangle =
\Delta^2 \delta_{jk},
\end{equation}
 where $\Delta^2 = \bar{\delta^2} - \bar{\delta}^2$.
Replacing in Eq.~(\ref{eq:Sigma}) and since $\langle n \rangle = (1/N_{i})
\sum_j^{N_i} n_j = S_0 /N_{i}$, we obtain
\begin{equation}
\Big(\sigma_1(S_0)\Big)^2 = N_{i} \Delta^2
\sim S_{0}.
\end{equation}
Comparing with Eq.~(\ref{eq:Sigma_exponent}) we obtain $\beta_{\rm
  rand} = 1/2$ for this uncorrelated case.

Let us assume that the correlation of the population increments
$\delta_j$, decays as a power-law of the distance between cells
indicating long-range scale-free correlations. Thus, asymptotically

\begin{equation}
\langle (\delta_j - \bar{\delta}) (\delta_k - \bar{\delta}) \rangle
\sim \frac{\Delta^2} {|\vec{x}_{j} - \vec{x}_{k}|^{\gamma}},
\end{equation}
 where
$\vec{x}_{j}$ denotes the position of the cell $j$ and $\gamma$ is the
correlation exponent (for $|\vec{x}_{j} - \vec{x}_{k}| \rightarrow 0$,
the correlations $\langle (\delta_j - \bar{\delta}) (\delta_k -
\bar{\delta}) \rangle$ tend to a constant). For large clusters, we can
approximate the double sum in Eq.~(\ref{eq:Sigma}) by an
integral. Then, assuming that the shape of the clusters can be
approximated by disks of radius $r_c$, for $\gamma<2$ we obtain
\begin{equation} (\sigma_1(S_0))^2 = \sum_{j,k}^{N_{i}}
  \frac{\Delta^2}{|\vec{x}_{j} - \vec{x}_{k}|^{\gamma}} \rightarrow
  \Delta^2 \frac{N_{i}}{a^2} \int^{r_c} \frac{r {\rm d}r {\rm
      d}\theta}{r^\gamma} \approx \frac{\Delta^2}{(2-\gamma)}
  \frac{N_{i}}{a^2} r_c^{-\gamma+2},
\end{equation} where $a^2$ is the area of each cell and $r_c$ the
radius of the cluster. Since $r_c \sim N_{i} a^2$, we finally obtain,
\begin{equation}
  \Big(\sigma_1(S_0)\Big)^2 \sim N_{i}^{2-\frac{\gamma}{2}}.
\end{equation}
Using $S_0=N_{i} \langle n \rangle$ and Eq.~(\ref{eq:Sigma_exponent})
we arrive at,
\begin{equation}
  \label{eq:betagamma} \beta = \frac{\gamma}{4}.
\end{equation}

Equation (\ref{eq:betagamma}) shows that Gibrat's Law is recovered
when the correlation of the population increments is a constant,
independent from the positions of the cells; that is when all the
populations cells are increased equally. In other words, if $\gamma =
0$, the standard deviation of the populations growth rates has no
dependence on the population size ($\beta = 0$), as stated by Gibrat's
law. The random case is obtained for $\gamma=d$, where $d=2$ is the
dimensionality of the substrate. In this case $d=2$ and $\beta_{\rm
  rand}=1/2$. For $\gamma>2$, the correlations become irrelevant and
we still find the uncorrelated case $\beta_{\rm rand}=1/2$. For
intermediate values $0<\gamma<2$ we obtain $0<\beta = \gamma/4 <1/2$.

\section{Random surrogate dataset}
\label{random}

In this section we elaborate on the randomization procedure used to
understand the role of correlations in population growth.

Figure~4C in the main text shows the standard deviation $\sigma(S_0)$
when the population of each cluster is randomized, breaking any
spatial correlation in population growth. For clusters with a large
population, $\sigma(S_0)$ follows a power-law with
exponent $\beta_{\rm rand}=1/2$, and for small $S_0$, $\sigma(S_0)$
presents deviations from the power-law function as seen in
Fig.~4C with smaller standard deviation than the
prediction of the random case. This deviation is caused by the fact
that the population of a cluster is bound to be positive: a cluster
with a small population $S_0$ cannot decrease its population by a
large number, since it would lead to negative values of $S_1$. This
produces an upper bound in fluctuations of the growth rate for small
$S_0$ and results in smaller values of $\sigma(S_0)$ than expected
(below the scaling with exponent $\beta_{\rm rand}=1/2$).

\begin{figure}
\includegraphics[width=8cm]{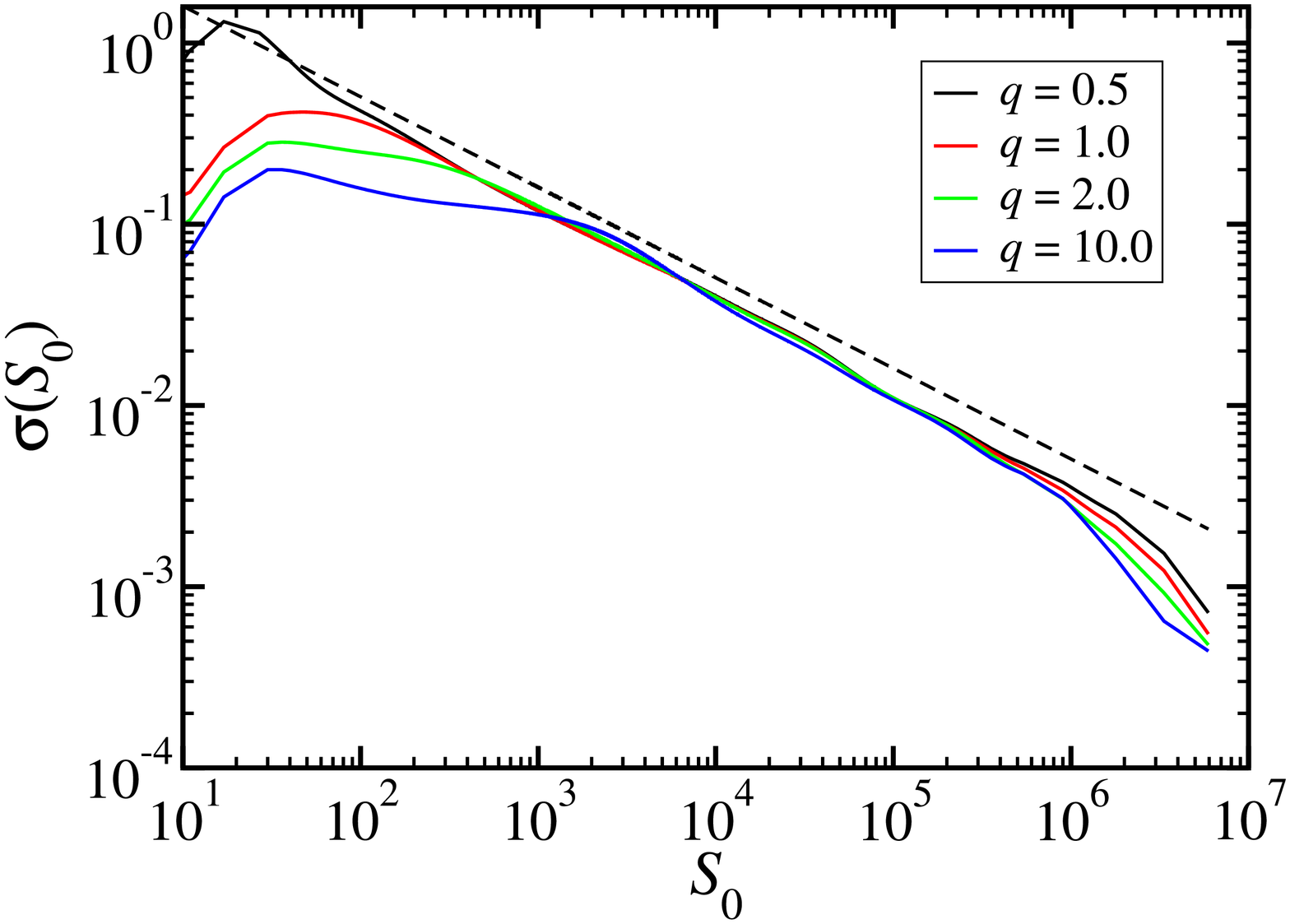}
\caption{ Standard deviation $\sigma(S_0)$ for the random data
    set as explained in the SI Section~\ref{random}. The results for
  $\sigma(S_0)$ are rescaled to collapse the power-law tails with
  exponent $\beta_{\rm rand}=1/2$ and to emphasize the deviations from
  this function for small values of $S_0$. The larger the parameter $q$, the
  larger the deviations from the power-law at lower $S_0$. In other
  words, the crossover to power-law tail appears at larger $S_0$ as
  $q$ increases.  }
\label{random_growth}
\end{figure}

To support this argument, we carry out simulations using the clusters
of GB, where the population $n_j(t_0)$ of each cell $j$ is replaced
with random numbers following an exponential distribution with
probability $P(n_j) \sim {\rm e}^{-n_{j}/n_{0}}$. The decay-constant, $n_{0}
= 150$, is extracted from the data of GB to mimic the original
distribution. This is done through a direct measure of $P(n_j)$ 
from the GB dataset and fitting the data using OLS regression analysis.
We obtain the population $n_j(t_1)=n_j(t_0) + \delta_j$ of cell $j$ at
time $t_1$ by picking random numbers for the population increments
$\delta_j$ following a uniform distribution in the range $-q*150 <
\delta_i < q*150$.
Here $q$ determines the variance of the increments.  Since the
population cannot be negative we impose the additional condition
$n_j(t_1)\ge0$.  Figure~\ref{random_growth} shows the results of the
standard deviation $\sigma(S_0)$ for four different $q$-values for
this uncorrelated model.  We find that the tail of $\sigma(S_0)$
reproduces the uncorrelated exponent $\beta_{\rm rand}=1/2$.  For
small $S_0$ we find that the standard deviation levels off to an
approximately constant value as in the surrogate data of
Fig~4C.  The crossover from an approximately constant
$\sigma(S_0)$ to a power-law moves to smaller values of the population
$S_0$ as the standard deviation in the $\delta_j$ is smaller (smaller
value of $q$).  Such behavior can be understood since the condition
$n_j^{(i)}(t_1)\ge0$ imposes a lower ``wall'' in the random walk
specified by $n_j^{(i)}(t_1)=n_j^{(i)}(t_0) + \delta_j$.  As the
initial population gets smaller, the walker ``feels'' the presence of
the wall and the fluctuations decrease accordingly, thus explaining
the deviations from the power-law with exponent $\beta_{\rm rand}=1/2$
for small population values.  Therefore, as the value of $q$
decreases, the small population plateau disappears as observed in
Fig.~\ref{random_growth}.

\section{A variation of the CCA}
\label{variation}

\begin{figure}
\centering {
\hbox{{\bf A} \resizebox{0.5\textwidth}{!}{\includegraphics{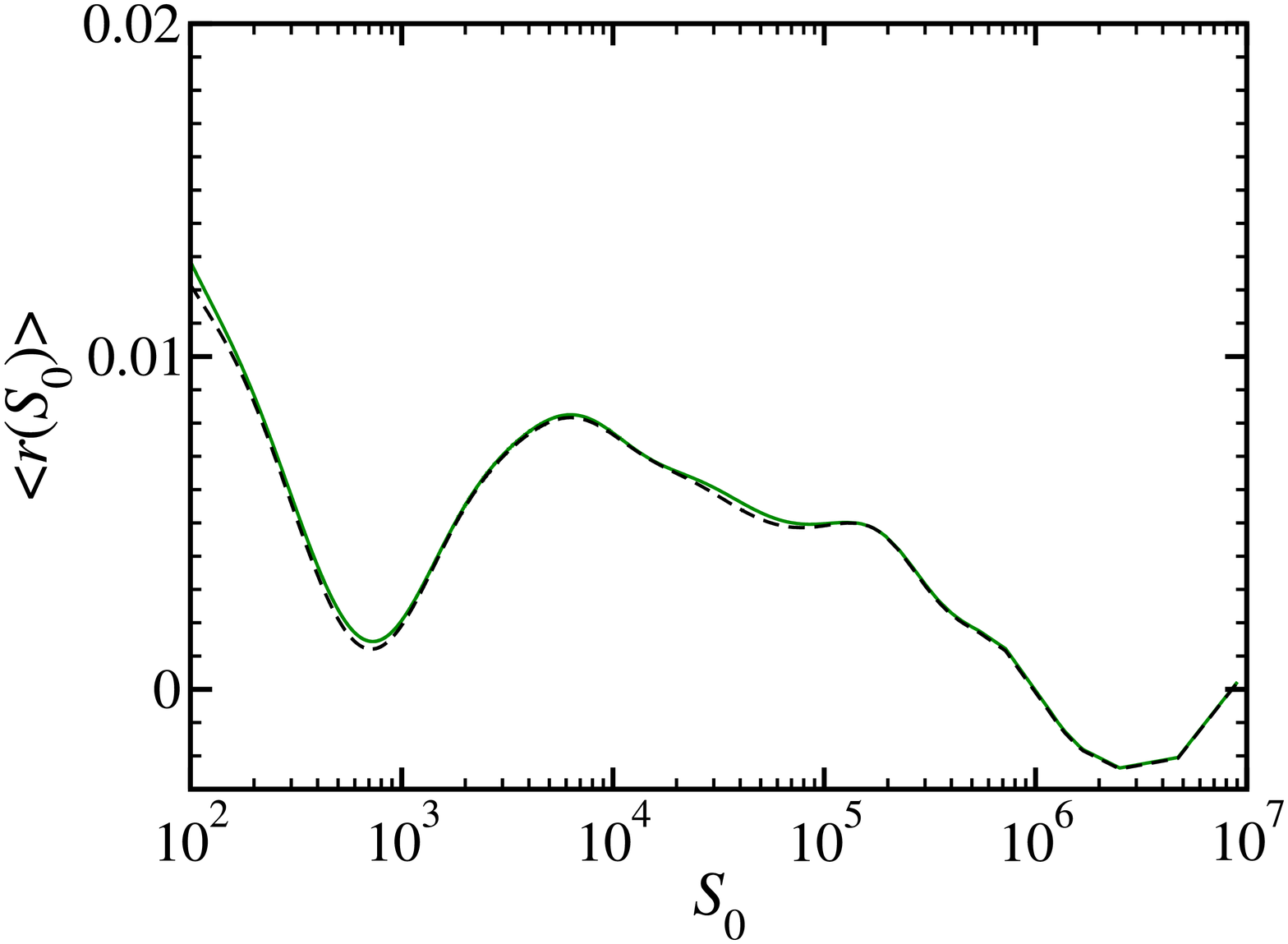}}
{\bf B} \resizebox{0.5\textwidth}{!}{\includegraphics{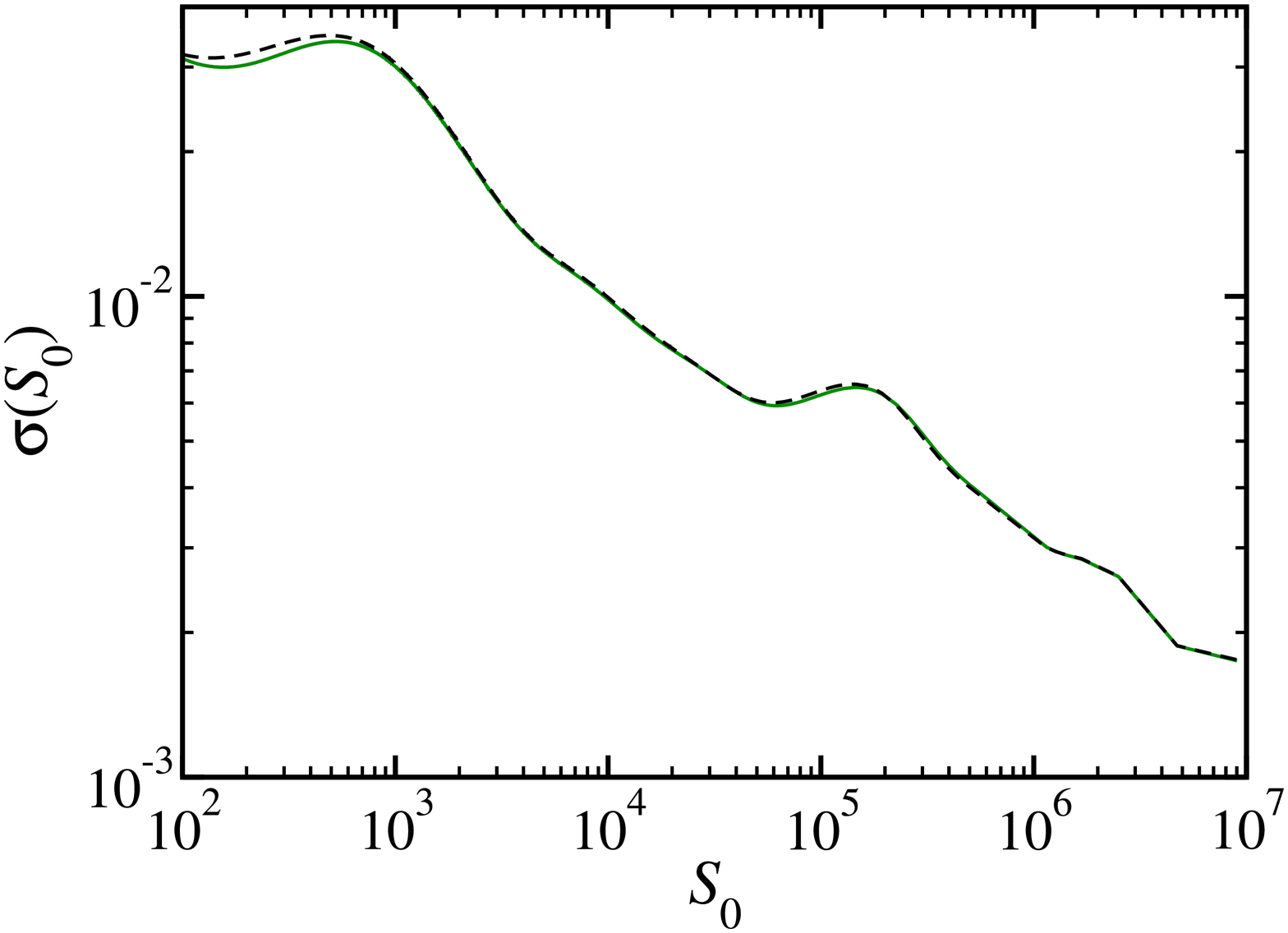}}
}}
\caption{Sensitivity of the results under a change in the stopping criteria in the CCA
 {\bf (A)} Average growth rate for GB with a population threshold of 5 (green line) and 20 (black dashed line)
 and {\bf (B)} standard deviation
  for GB with a population threshold of 5 (green line) and 20 (black dashed line).
   For clarity we do not show the confidence bands. }
\label{threshold}
\end{figure}

In this section we study a variation of the CCA. In the main text we stop growing a cluster when the population of all boundary cells have unpopulated, that is, have population exactly 0. In other words, clusters are composed by cell with population strictly greater than 0.
 It is important to analyze whether this stopping criteria can be relaxed to including cell which have a population larger that a given threshold. In Fig.~\ref{threshold}A and Fig.~\ref{threshold}B we show the results for the population growth rate and standard deviation, respectively, in GB when the cell size is 2.2km-by-2.2km (as in the main text) but including cells with a population strictly larger than 5 and 20. 

Although for small population clusters we observe a slight variation in the growth rate and in the standard deviation, the results show that the thresholds do not influence the global statistics when compared to the plots in the main text. 

\clearpage

\end{document}